# Chapter Title. Fishery resources management[1]

**Authors**

Hidekazu Yoshioka[1, *], Motoh Tsujimura[2], and Yumi Yoshioka[3]

[1]Japan Advanced Institute of Science and Technology, 1-1 Asahidai, Nomi, Ishikawa 923-1292, Japan. ORCID: 0000-0002-5293-3246

[2]Doshisha University, Karasuma-Higashi-iru, Imadegawa-dori, Kamigyo-ku, Kyoto 602-8580, Japan. ORCID: 0000-0001-6975-9304

[3]Gifu University, Yanagido 1-1, Gifu, Gifu 501-1193, Japan. ORCID: 0000-0002-0855-699X

*Corresponding author: yoshih@jaist.ac.jp, +81-761-51-1745

**Abstract**

We consider management of the fish species *Plecoglossus altivelis altivelis*, a major inland fishery resource in Japan playing important roles from economic, cultural, and recreational viewpoints. We firstly summarize the collected body weight data of the fish in the Hii River, Japan since 2016. The two kinds of data are available in each year with few exceptions: the historical data during summer and autumn collected with the help of an angular and the annual distribution data at the Toami (casting net) competition where we could obtain the data from many anglers during two hours in one day. We fit deterministic and uncertain logistic growth models to the data in each year and discuss their performance. The fitted uncertain logistic growth model is applied to an optimal harvesting problem of the fish subject to a sustainability concern and model distortion. Several numerical schemes for solving the problem are examined and compared both theoretically and numerically.

(154 words)

**Keyword**

Inland fishery resources; Harvesting; Body weight; Historical data; Distribution data; Toami competition; *Plecoglossus altivelis altivelis*; Uncertain logistic growth model; Optimal control; Finite difference methods

---

[1] This manuscript is a preprint version of a chapter manuscript for a book.

## 6.1 Introduction

### 6.1.1 Background

Fisheries in inland water bodies, such as lakes and seas, are called inland fisheries and have smaller scales than marine fisheries (only 10% of total fish catch), while they supply food and support human well-being (Cooke et al., 2016). According to Ainsworth et al. (2023), inland waters support the livelihoods of up to 820 million people on Earth, and 90% of inland fisheries catch is consumed in the developing world. Nguyen et al. (2023) suggested the need for conducting long-term monitoring of river ecosystems and preparing fisheries law to preserve biodiversity and ecosystem services provided by inland fisheries and found that inland fisheries provide important provisioning and cultural services for residents. Because inland fisheries rely on fish as biological resources living in water environments, climate change would critically reduce their population by increasing the frequency of unpredictable dry seasons or drought events (Nyboer et al., 2022). In addition to climate change, habitat degradation, pollution, overexploitation, and species invasion negatively affect inland fisheries (Stokes et al., 2021). Inland fisheries are often controlled by local stakeholders, such as practitioners, managers, resource biologists, and stewardship officers working within administrations. A unified scheme to collectively make their actions affects sustainability of fishery resources (Cooke et al., 2021). Improvements and clarification of fishery rules, as well as more deliberation, have been suggested to be crucial for sustaining fisheries without environmental degradation and overexploitation (Stratoudakis et al., 2023). Surveying, modeling, and analysis of inland fisheries are therefore indispensable for achieving sustainability.

There are wide ranges of mathematical modeling studies on inland fisheries. Ahrens et al. (2019) constructed a population dynamics model of size-structured fishes and explored harvesting strategies to meet multiple management objectives. Voss and Quaas (2022) formulated a discrete-time optimization model of cod fishing and studied regime switching of the fish population between "good" and "bad" states. Dao et al. (2023) investigated a static stability problem of bioeconomic model for regulating fishery resource harvesting through a resource-price feedback mechanism. Mitra et al. (2023) statistically analyzed the relationship between open water availability and efficiency of tilapia farming ponds and suggested their environmental management schemes depending on the local water resource availability. Kane et al. (2022) proposed a statistical model to estimate the fishery resource population in an inland water body as a function of its surface area, which was later applied to estimate the efforts of boats and bank anglers in water bodies (Kane et al., 2023). Tucker et al. (2024) applied Bayesian method to creel-survey data from multiple sources to obtain a common estimate of angler efforts.

Biological and ecological dynamics are based on complex and nonlinear interactions among many processes; hence, their complete mathematical description is extraordinarily difficult, if not impossible, which motivates the use of a mathematical model that accounts for misspecification of coefficients and/or parameter values. This issue has been discussed for biological population dynamics using sigmoid curves (Simpson et al., 2022), chemical reactions in systems biology (Murphy et al., 2024), spatial population dynamics models focusing on fisheries management (Berger et al., 2021), and species interactions in ecological networks (Adams et al., 2020). However, for inland fisheries management, a consistent approach starting from field surveys, formulation of a mathematical model, its theoretical analysis, development of a reliable numerical method to discretize the model, and their applications are insufficient. In this chapter, we present such an example based on optimal control theory.

**6.1.2 Objective and contribution**

The aim of this chapter is to summarize fisheries management data in our study site, the Hii River in Japan, particularly biological growth data of the fish *Plecoglossus altivelis altivelis* (*P. altivelis*). This fish species is a major inland fishery resource in the country as already reviewed in **Chapter 5**. The data imply that individual differences in body weight are not negligible and should be incorporated into the resource management of fish. The core mathematical approach here assumes that the individual difference is caused by the difference of the maximum body weight and that everyone grows following a logistic growth curve that has different maximum body weights as uncertainties, namely difference among individuals.

As an application of the biological growth model to fisheries management, we consider a cost-efficient harvesting problem of *P. altivelis* during a single fishing season, with or without sustainability concerns. Sustainability here means that the fish population should not be overexploited during the fishing season so that the population does not become extinct. We follow the approach of Yoshioka (2024a) in a more analytically tractable form. From a different standpoint, the state variable to be controlled is not a fish population in a target water body but a harvested population from it; the former is difficult to observe or estimate in the real world, while the latter is not. The population dynamics of the fish is assumed to follow an ordinary differential equation (ODE) subject to continuous-time harvesting as a control variable. We then formulate an optimal control problem in a finite horizon whose objective function has the three terms: cumulative harvesting utility, cumulative harvesting cost, and a terminal profit if there is a sustainability concern. The growth curve along with its misspecification is incorporated into the harvesting utility in the form of biomass.

We argue that the optimality equation associated with the control problem, the Hamilton–Jacobi–Bellman (HJB) equation, is a partial differential equation (PDE) in a nonlinear hyperbolic form as in the literature (e.g., Fleming and Soner (2006)). A remarkable property of our HJB equation in this chapter is tractability that solution profiles can be studied analytically (i.e., monotonically increasing under certain conditions), and further an easily implementable explicit, semi-implicit, and implicit numerical schemes be applied to its numerical discretization. In general, a fully implicit numerical method for a PDE like an HJB equation requires solving some matrix system whose inversion (e.g., Forsyth and Vetzal, (2012)) will be time-consuming. By contrast, our HJB equation, due to its unique mathematical structure (i.e., functional form of the Hamiltonian) can be implicitly handled without solving any matrix systems. The implicit nature also contributes to computational stability.

**6.2 Fisheries data analysis**

**6.2.1 Inland fishery cooperatives in Japan**

We explain some key facts in inland fisheries management in Japan. These are not directly related to mathematical modeling and particularly the control in this chapter; however, they will be helpful for better understanding our study background.

The following data are according to Fisheries Agency (2021). Total number of union members of inland fishery cooperatives in Japan has been decreasing from about 550 thousand in 1980's to about 270 thousand in 2020's. It has also been predicted that the total number of union members will decrease in the future as well and will become less than 200 thousand, and possibly approaches to around 100 thousand in 2030's. Total number of inland fishery cooperatives is gradually decreasing on average from 888 in 2000 to 802 in 2019. In 2015, more than 70.6% of union members are older than the age 60; in the same year, 36.3% of people in Japan were older than sixty. Union members in inland fishery cooperatives in the country are more weighted to older ages, and thus they are facing a strong wave of aging.

The inland fishery cooperative that authorizes fishery resources in the Hii River in Japan, the Hii River Fishery Cooperative (HRFC), is one of the fishery cooperatives that we have been working together. Our collaboration started in 2015. Total number of union members of HRFC is 835 in 2021, 805 in 2022, and 787 in 2023 (HRFC, 2021; HRFC, 2022; HRFC, 2023) and is therefore gradually decreasing in recent years. The age distribution of union members in the HRFC is not known, but in our experience many of them are retirees. We therefore consider that the HRFC is also facing a wave of aging. According to Matsuda et al. (2021), 38% and 71% of inland fishery cooperatives in Japan have less than 100 and 300 union members, respectively. The HRFC is thus considered to be in the superior 29% of all inland fisher cooperatives in Japan ranked in the total number of union members. Therefore, the HRFC is suggested to be a medium to large inland fishery cooperative in the country, and they may still have some energy left for collaborating with researchers like us. However, the aging and decline is a critical issue to continue their duties to authorize fishery resources and their surrounding river environment in the future.

### 6.2.2 *P. altivelis* and classical logistic model

The target fish *P. altivelis* is a diadromous fish species widely distributed in Northeast Asia along the Sea of Japan (East Sea) and East China Sea. Its annual life history allows for discriminating two ecological forms, which are the amphidromous form migrating between river and sea and landlocked form migrating between river and lake (Tsukamoto and Uchida 1992). An important fact in the context of the model in this chapter is that juvenile fishes migrate upstream from sea or lake toward river during spring, grow and mature in river during summer, and then die after spawning during autumn (**Figure 6.1**). The fish size of upstream migration is positively correlated to the river discharge (Mouri et al. 2010), and the timing of upstream migration becomes earlier as the river water temperature during spring is relatively higher (Suzuki et al., 2004). In addition, upstream migration from estuary toward river occurs more often with higher temperature and discharge (Yoon et al. 2016). The downstream migration is also triggered by a temperature threshold, and the spawning has been reported to start when the water temperature falls below 20 °C (Tran et al. 2017). In general, the harvesting season of the fish is in mid-summer to the coming autumn before or around the spawning. The authors have recently been continuously investigating biological growth of the fish *P. altivelis*. The fish has been a major inland fishery resource in Japan continuously playing vital roles in driving local economy (Yoshimura et al., 2021) and shaping regional culture (Hori et al., 2022).

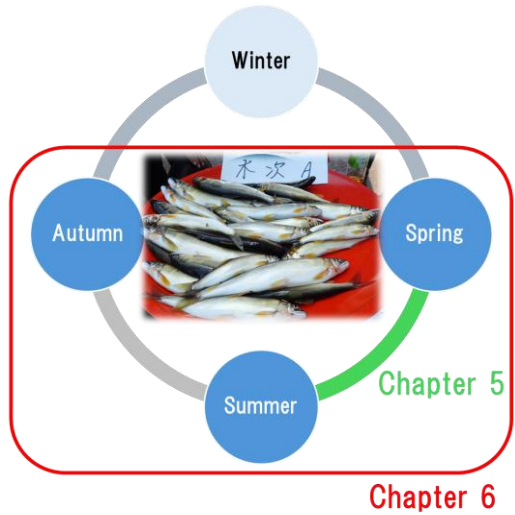

**Figure 6.1.** Illustration explaining the life history of the fish *P. altivelis*.

Most anglers catch fish for a recreational. In many cases, they use the Tomozuri angling (fishing decoy: e.g., Aino et al. (2015)) or Toami (casting net: e.g., Otake et al. (2002)). Fishing of *P. altivelis* in Japan has been authorized by local inland fishery cooperatives with the fishing season from June to July (summer) to October to November (autumn) with differences depending on the region. Inland fisheries of midstream to upstream reaches of the Hii River have been authorized by the HRFC as explained earlier. The fishing season of *P. altivelis* in the Hii River has been set as from July 1 to December 31 in each year (HRFC, 2023), but almost no anglers are found during December because the spawning season of the fish is around October in the study site, after which the adult fishes die out in general. In each spring it has been reported by residents that juveniles of *P. altivelis* migrated from the Sea of Japan toward the midstream of the Hii River; often migrating juveniles were observed in weirs installed in the river. Furthermore, the HRFC released several thousands of *P. altivelis* individuals into the Hii River to sustain the fish population. Both natural and released *P. altivelis* coexist in the Hii River. Both amphidromous and landlocked fish were present at the study site. However, their ratios are not known.

The body weight data of *P. altivelis* in the Hii River system has been collected with the help of a union member of the HRFC each year since 2016. During each fishing season, this person arrived at the Hii River almost once in a couple of days and caught 27.3 individuals of *P. altivelis* on average according to the data from 2018 to 2023. After each arrival at the river, this person counted the total number of caught fish and measured their average body weight, but not their individual weights, due to limitations of labor.

We have another set of data on fish body weight, which is the data collected at the Toami competition, an annual competition held by the HRFC. More than ten pairs were registered at each competition. The pair with the largest fish catches during the competition (typically between 10:00 am to 12:00 am on a summer day in the growing season of the fish) was awarded by the HRFC. In each competition approximately 200 to 300 individuals were fished, and the body-weight data of *P. altivelis* in 2017, 2018, 2019, and 2023 were successfully collected. The competition was not held in 2020, 2021, and 2022 because of the Covid-19 at the end of 2019. We also obtained the total fish catch, average body weight, and maximum and minimum body weights at the competition in 2016, but without data on individuals.

Consequently, we have the historical average body weight data of *P. altivelis* in the Hii River since 2016 and individuals' body weights in one day for a few years since 2017, with which a growth model of the fish in the river can be found. A common limitation of these data is that they are available only in summer and autumn but not in spring and winter, and hence do not cover the larval and early juvenile periods of the fish. Therefore, we estimated the growth dynamics of fish in that period by extrapolation.

Below, we analyze each dataset and fit the mathematical models against the data. In each figure presented below with the unit of time being day, the beginning of May 1 is referred to as time 0 (day) because it has been suggested that natural *P. altivelis* are migrating from the sea to the midstream Hii River and farmed fish are released around this day each year. The day July 1 at the beginning of the fishing season of the fish starts from the time 61 (day).

**Figure 6.2** shows the collected (averaged) body weight since 2016, showing that the observed body weights fluctuated over time in each year. **Figure 6.2** also shows the best least-squares fit by the logistic growth curve as the simplest sigmoidal growth curve to explain biological growth of fishes (Yoshioka and Yaegashi, 2016):

$$W_t = \frac{W_{\max}}{1 + \left(\frac{W_{\max}}{W_0} - 1\right)\exp(-rt)} \quad \text{for} \quad t \geq 0 \tag{6.1}$$

with the initial body mass $W_0 > 0$, maximum body weight $W_{\max} (> W_0)$, and growth rate $r > 0$. The time zero can be decided by a modeler. This body weight in (6.1) satisfies the so-called logistic equation

$$\frac{dW_t}{dt} = rW_t\left(1 - \frac{W_t}{W_{\max}}\right) \tag{6.2}$$

for any $t > 0$ with the initial condition $W_0 > 0$. This model describes a biological growth with the net growth rate given by $r\left(1 - \frac{W_t}{W_{\max}}\right)$, which is increasing and decreasing with respect to the growth rate $r$ and maximum body weight $W_{\max}$, respectively. Intuitively, this parameter dependence shows that an individual with a higher growth rate or maximum body weight grows faster. The fitted parameter values of the logistic growth curve are summarized in **Table 6.1**. The logistic model serves reasonably well for fitting the mean growth of *P. altivelis* in the Hii River; however, the size spectrum, namely the individual difference in body size, is not negligible, but cannot be accounted for by the logistic growth curve (6.1) in principle.

**Figure 6.3** shows the collected individual body weights at the Toami competition each year, supporting the existence of individual differences. **Table 6.2** summarizes the empirical statistics for each competition. The average body weight is smallest in 2023, which is less than 2 (g) to 3 (g) from those in the earlier years. This is partly due to using the one-week earlier event data in 2023 than in the other years.

Indeed, if we employ a crude assumption that the upstream migration of *P. altivelis* occurs on the same date each year, then the earlier competition data imply a larger probability of catching smaller fish because the fish has a one-year life history. Note that the body weight data collected in this chapter inherently have observation errors in that the correct time at which each fish was harvested is not available but has an error of up to 24 (hours). However, these errors do not critically affect the discussion in this chapter. This is because the relative error will be 0.2 (g) to 1.0 (g) according to the identified logistic growth curves and the minimum and maximum errors reached in late October and early July, respectively.

### 6.2.3 Uncertain logistic model

The existence of individual differences in the data for *P. altivelis* suggests that the growth curve for this fish should not be deterministic but rather probabilistic, such that different individuals have different growth histories. Several mathematical models have been proposed for growth curves of biological species. They include but are not limited to the nonlinear stochastic differential equation driven by Brownian noises that generate irregular growth trajectories (Campillay-Llanos et al., 2024; Jacinto et al., 2022), an open-ended logistic growth model with a probabilistic maximum body weight that generates smoother growth trajectories (Yoshioka et al., 2019), a Markovian kernel model whose parameters grow randomly as time elapses with the constraint that the growth dynamics satisfy Kolmogorov's equation (Johne et al., 2023), and an uncertain-parameter growth curve model that assumes the common form of the growth curve among individuals but with different parameter values among them (Bevia et al., 2023). We employ the model of Bevia et al. (2023), which was applied to the growth data of *P. altivelis* for 2021 and 2022 in an earlier study (Yoshioka, 2024a). Similar modeling methodologies based on random parameter values have been applied to the other phenomena like the disease propagation (Jornet, 2022) and shock wave generation (Dorini et al., 2023)

The model, which we call the uncertain logistic model, is a generalization of the logistic growth curve (6.1) but with uncertain maximum body weight $W_{\max}$ as a distributed parameter in a fixed range $\Omega_W = (\underline{W}, \overline{W})$ with $0 < \underline{W} < \overline{W} < +\infty$. The realization of $W_{\max}$ is assumed to follow a probability distribution having the domain $\Omega_W$. In this view, the logistic growth curve (6.1) is a realization of the uncertain logistic model. If we denote the probability measure to generate $W_{\max} \in \Omega_W$ by $P(\mathrm{d}W_{\max})$ ($\int_{\Omega_W} P(\mathrm{d}W_{\max}) = 1$), then the average body weight $W_t^{(1)}$ of individuals at time $t > 0$ is obtained as

$$W_t^{(1)} = \int_{\Omega_W} \frac{W_{\max}}{1 + \left(\frac{W_{\max}}{W_0} - 1\right)\exp(-rt)} P(\mathrm{d}W_{\max}) \quad \text{for} \quad t \geq 0. \tag{6.3}$$

Other statistics can be obtained in the same manner. For example, the $m$-th moment of the body weight $W_t^{(m)}$ ($m = 1, 2, 3, ...$) is

$$W_t^{(m)} = \int_{\Omega_W} \left( \frac{W_{\max}}{1+\left(\frac{W_{\max}}{W_0}-1\right)\exp(-rt)} \right)^m P(dW_{\max}) \quad \text{for } t \geq 0. \tag{6.4}$$

With this in mind, we obtain statistics having a higher order than the average, such as the variance $W_t^{(2)} - \left(W_t^{(1)}\right)^2$.

One may also consider a model where the parameters $r, W_0$ are distributed as well. This is theoretically possible but leads to a more complicated model with a more complicated integration. Our assumption that the maximum body weight is uncertain is in accordance with the intuition that both large and small individuals of *P. altivelis* are found in the river during the late growth season (i.e., around October). The parameter dependence of the logistic growth curve discussed above suggests that a larger maximum body weight implies a higher net growth rate. Therefore, our assumption leads to a mathematical model in which an individual with a larger maximum body weight eventually becomes larger at a faster speed.

To compare the obtained growth data in each year using the uncertain logistic model, we need to identify the parameter values of $r, W_0$, the range $\Omega_W$ with the upper $\overline{W}$ and lower bounds $\underline{W}$, and the probability measure $P$. We continue with the following parameter identification strategy that can better exploit availability of the body weight distribution in 2017-2019 and 2023, presented in **Figure 6.3**. First, we select a year in which a Toami competition was held. Then, we let the subscript "e" and "m" represent empirical and modeled values, respectively. We try to choose the parameters $r, W_0, \underline{W}, \overline{W}$ and probability density $P$, such that the following error metric Er is minimized:

$$\text{Er} = \left( \frac{\text{Ave}_e - \text{Ave}_m}{\text{Ave}_e} \right)^2 + \left( \frac{\text{Std}_e - \text{Std}_m}{\text{Std}_e} \right)^2. \tag{6.5}$$

The error Er is a sum of the squares of the relative errors of average and standard deviation, suggesting that the method we use is a naïve moment-matching method. After identifying the parameters and probability density minimizing the error metric Er, we check whether the identified model captures the other statistical characteristics, such as the positivity of skewness, and the historical data. One may incorporate a term considering the relative error of higher-order statistics, such as a skewness, in the error metric (6.5). However, our preliminary investigations suggested that such a procedure fits only the data of the Toami competition, but not necessarily the historical data well.

We need to specify the functional form of the probability measure $P$ to implement the parameter identification procedure. According to the empirical data collected at each Toami competition, the body weight distribution of the fish has a positive skewness (**Table 6.2**). This empirical finding combined with the increasing property of the logistic growth curve on the maximum body weight $W_{\max}$ suggests that the probability measure $P$ admits a probability density function $p$ (PDF) having a positive skewness. A convenient choice of the PDF satisfying this property is the beta distribution:

$$p(w) = C(w - \underline{W})^{a-1}(\overline{W} - w)^{b-1}, \quad w \in \Omega_W \tag{6.6}$$

with parameters $a, b > 0$. Here, $C > 0$ is the normalization constant uniquely determined to give $\int_{\Omega_W} p(w)\mathrm{d}w = 1$. As shown in **Figure 6.4**, the beta distribution is positively (skewed resp., negatively skewed) if $a > b$ (resp., $a < b$). At this stage, parameters to be identified are as follows: $r, W_0, \underline{W}, \overline{W}, a, b$. We further reduce the total number of parameters to be identified, by assuming that either $W_0 = 10$ (g) for all years according to the personal communication with the HRFC that the body weight of the released *P. altivelis* is around this value, or $W_0$ by the logistic growth curve in **Table 6.1**. Consequently, the parameters to be identified are $r, \underline{W}, \overline{W}, a, b$. We employ a trial-and-error approach where such that the following set of parameters are explored to find the minimizer of the error metric (6.5):

$$(r, \underline{W}, \overline{W}, a, b) = (0.020 + 0.001i \ (1/\text{day}), j \ (\text{g}), k \ (\text{g}), 0.25l, 0.25m) \tag{6.7}$$

with $0 \leq i \leq 40$, $1 \leq j \leq 50$, $1 \leq j \leq k \leq 300$, $1 \leq l, m \leq 40$. The integration with respect to $P$ is conducted with the midpoint rule with 1,000 evaluation points.

**Tables 6.3 and 6.4** summarize the identified parameter values along with the error metric (6.5) for each year and each initial condition. **Figure 6.5** shows the empirical data and the theoretical results with the identified model in each year with $W_0 = 10$ (g). **Figure 6.6** shows the results with $W_0$ based on **Table 6.1**. **Figure 6.7** compares the empirical and theoretical results for the Toami competition for each year. **Tables 6.3 and 6.4** show that in each case of $W_0$, the error metric Er is of the order of $10^{-5}$. This value is sufficiently small for the average and standard deviation; both statistics coincide up to three digits between the empirical and theoretical results in both cases. **Figures 6.5 and 6.6** show that all the empirical results are between the theoretical maximum and minimum growth curves, even though historical growth data were not used to identify the model. This finding supports the validity of the identified models. A comparison of these figures implies that the theoretical maximum and minimum growth curves are sensitive to $W_0$. This is due to a limitation of the proposed identification method, in which only the distribution data, that is, one-time data, is used to find the best model. In **Figures 6.5 and 6.6**, the empirical data are scattered around the curves of the average and those of the average plus standard deviation, reflecting the positive skewness of the identified model.

Skewness, which was not included in the error metric (6.5), is positive in all the fitted models with underestimations except for the year 2023 with $W_0 = 10$ (g) and the year 2017 with the yearly specific $W_0$. The misfit of skewness is the most significant in 2018 with $W_0 = 10$ (g), where the skewness of the fitted model is more than ten time smaller than the empirical one. The misfit is smaller in the models of 2017 and are smaller than those of 2018 with $W_0 = 10$ (g), where the skewness of the fitted model is more than twice smaller than the empirical one. These misfits appear in **Figure 6.7** as less skewed profiles of the theoretical PDFs of body weight than the empirical ones. The fitted PDFs in 2017, 2018, 2019 are unimodal and suggested to be continuous, while those in 2023 are decreasing and discontinuous because of $\alpha = 1$.

This difference is caused by the deviation of the data for 2023 from those of the other years, as shown in **Figure 6.3**.

**Table 6.1.** Summary of the fitted model parameter values of the logistic growth curve for each year.

| Year | 2016 | 2017 | 2018 | 2019 | 2020 | 2021 | 2022 | 2023 |
|---|---|---|---|---|---|---|---|---|
| $W_0$ (g) | 12.9 | 9.8 | 8.5 | 8.2 | 6.5 | 7.1 | 17.0 | 20.5 |
| $W_{max}$ (g) | 104.7 | 91.0 | 92.8 | 102.9 | 78.2 | 107.1 | 105.7 | 83.2 |
| $r$ (1/day) | 0.0241 | 0.0315 | 0.0298 | 0.0297 | 0.0383 | 0.0371 | 0.0223 | 0.0272 |

**Table 6.2.** Summary of the empirical statistics of body weight at each competition: N.A. means that the data are not available.

| Year | 2016 | 2017 | 2018 | 2019 | 2023 |
|---|---|---|---|---|---|
| Date | August 7 | August 6 | August 5 | August 4 | July 30 |
| Day (Day 0 is May 1) | 98 | 97 | 96 | 95 | 90 |
| Total fish catch | 207 | 234 | 189 | 227 | 297 |
| Average (g) | 55.2 | 55.6 | 57.3 | 56.4 | 52.2 |
| Standard deviation (g) | N.A. | 19.1 | 18.5 | 18.2 | 21.0 |
| Skewness (-) | N.A. | 0.8 | 1.2 | 0.9 | 1.4 |
| Median (g) | N.A. | 52.8 | 54.5 | 54.0 | 46.5 |
| Maximum (g) | 120.5 | 132.0 | 152.0 | 119.5 | 163.0 |
| Minimum (g) | 38 | 20.5 | 16.0 | 20.0 | 11.0 |

**Table 6.3.** Summary of the fitted model parameter values of the uncertain logistic model in each year with $W_0 = 10$ (g). The empirical and fitted statistics are also presented.

| Year | 2017 | 2018 | 2019 | 2023 |
|---|---|---|---|---|
| $\underline{W}$ (g) | 7 | 9 | 2 | 29 |
| $\overline{W}$ (g) | 177 | 147 | 151 | 293 |
| $\alpha$ (-) | 4 | 3 | 4.75 | 1 |
| $\beta$ (-) | 9.5 | 4.5 | 7.75 | 9.75 |
| $r$ (1/day) | 0.053 | 0.041 | 0.052 | 0.059 |
| Er (-) | 2.77E-05 | 1.36E-05 | 4.47E-05 | 5.33E-05 |
| Ave (empirical) | 55.6 | 57.3 | 56.4 | 52.2 |
| Ave (theoretical) | 55.6 | 57.3 | 56.4 | 52.2 |
| Std (empirical) | 19.1 | 18.5 | 18.2 | 21.0 |
| Std (theoretical) | 19.1 | 18.5 | 18.2 | 21.0 |
| Skew (empirical) | 0.77 | 1.15 | 0.95 | 1.42 |
| Skew (theoretical) | 0.38 | 0.08 | 0.18 | 1.43 |

**Table 6.4.** Summary of the fitted model parameter values of the uncertain logistic model in each year with $W_0$ in **Table 6.1**. The empirical and fitted statistics are presented as well.

| Year | 2017 | 2018 | 2019 | 2023 |
|---|---|---|---|---|
| $\underline{W}$ (g) | 24 | 24 | 8 | 24 |
| $\overline{W}$ (g) | 187 | 200 | 169 | 123 |
| $\alpha$ (-) | 2 | 1.75 | 4.5 | 1 |
| $\beta$ (-) | 8.25 | 5 | 10.25 | 2.5 |
| $r$ (1/day) | 0.075 | 0.038 | 0.066 | 0.079 |
| Er (-) | 3.34E-05 | 4.38E-05 | 6.43E-05 | 4.37E-05 |
| Ave (empirical) | 55.6 | 57.3 | 56.4 | 52.2 |
| Ave (theoretical) | 55.6 | 57.3 | 56.4 | 52.2 |
| Std (empirical) | 19.1 | 18.5 | 18.2 | 21.0 |
| Std (theoretical) | 19.1 | 18.5 | 18.2 | 21.0 |
| Skew (empirical) | 0.77 | 1.15 | 0.95 | 1.42 |
| Skew (theoretical) | 0.84 | 0.40 | 0.38 | 0.73 |

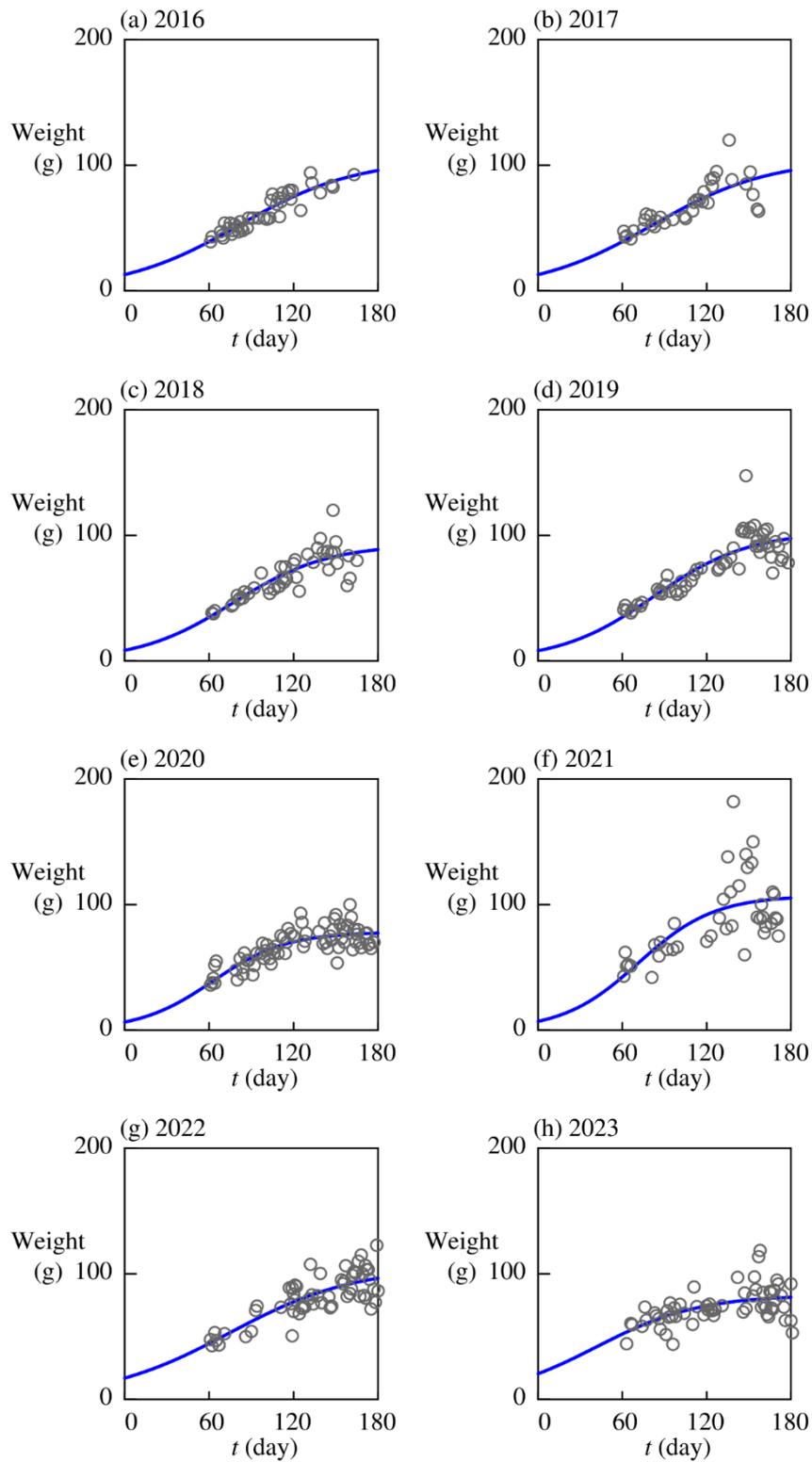

**Figure 6.2.** Empirical body weight data (circles) and fitted logistic growth (curves) for each year (the time 0 (day) is the beginning of May 1).

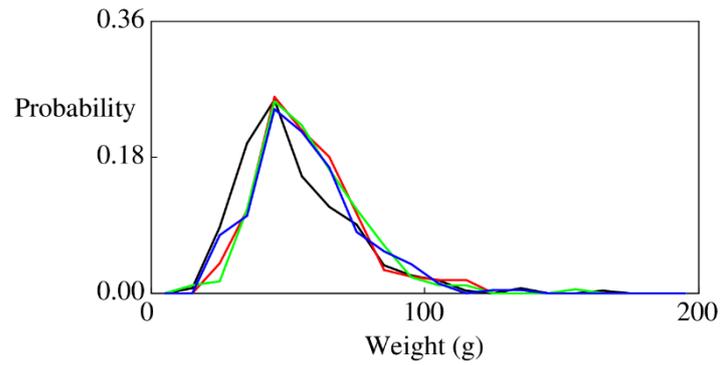

**Figure 6.3.** Individual body weights collected at the Toami competition in 2017 (blue), 2018 (green), 2019 (red), and 2023 (black).

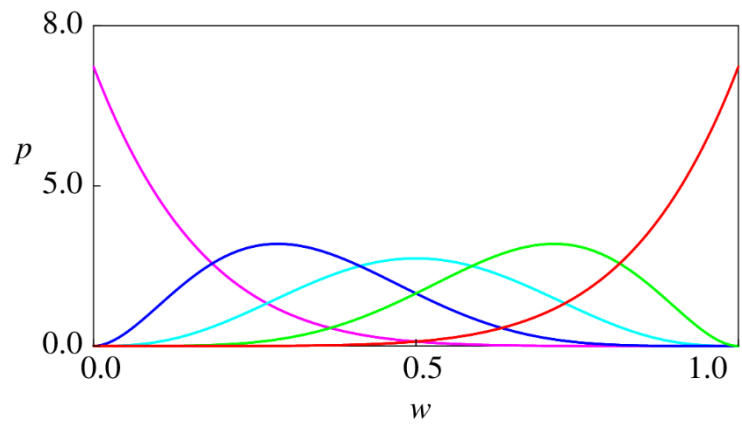

**Figure 6.4.** Beta distributions $p = p(w)$ for different parameter values of $(a,b)$ where we have used the normalization $\underline{W} = 0$ and $\bar{W} = 1$: $(a,b) = (7,1)$ (Magenta), $(a,b) = (6,3)$ (Blue), $(a,b) = (4,4)$ (Light blue), $(a,b) = (3,6)$ (Green), and $(a,b) = (1,7)$ (Red).

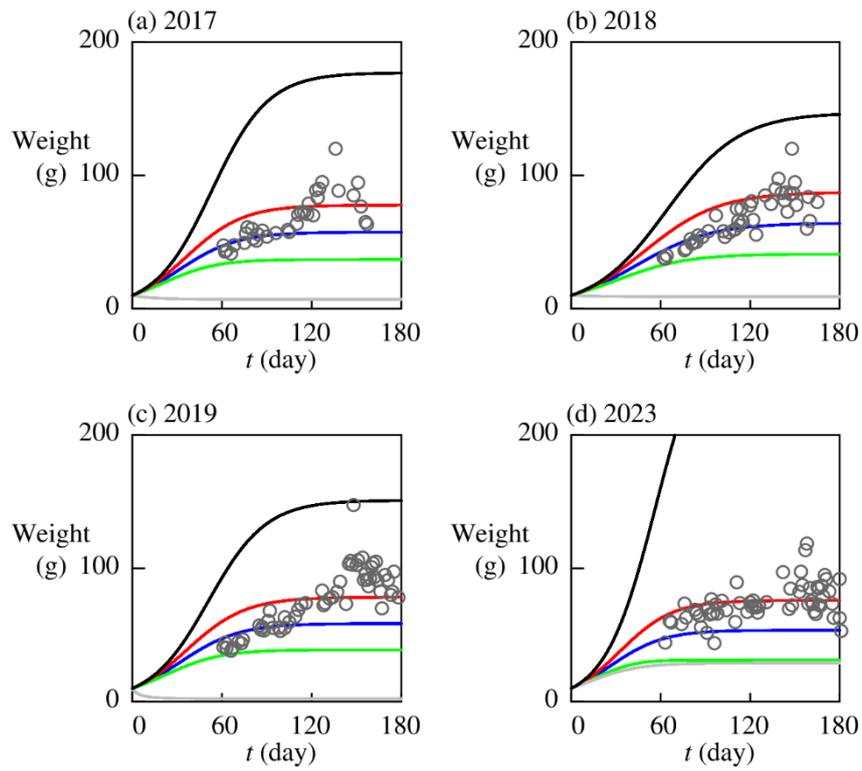

**Figure 6.5.** Empirical body weight data (circles) and the fitted uncertain logistic model (curves) with $W_0 = 10$ (g) (the time 0 (day) is the beginning of May 1): Ave (blue), Ave＋Std (red), Ave－Std (green), theoretical maximum (black), theoretical minimum (gray).

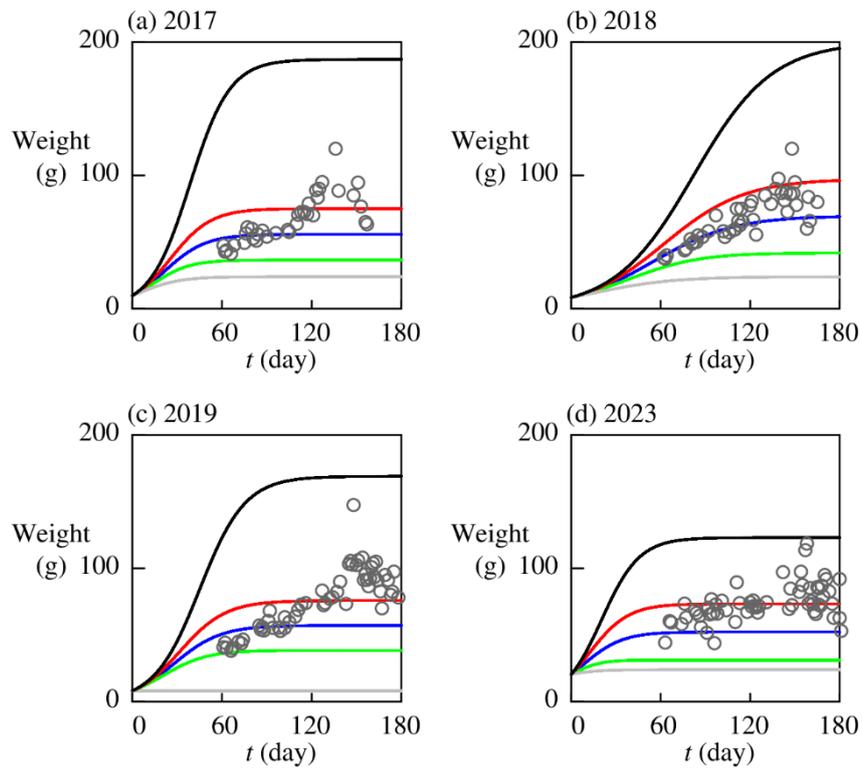

**Figure 6.6.** Empirical body weight data (circles) and the fitted uncertain logistic model (curves) with $W_0$ in **Table 6.1** (the time 0 (day) is the beginning of May 1): Ave (blue), Ave＋Std (red), Ave－Std (green), theoretical maximum (black), theoretical minimum (gray).

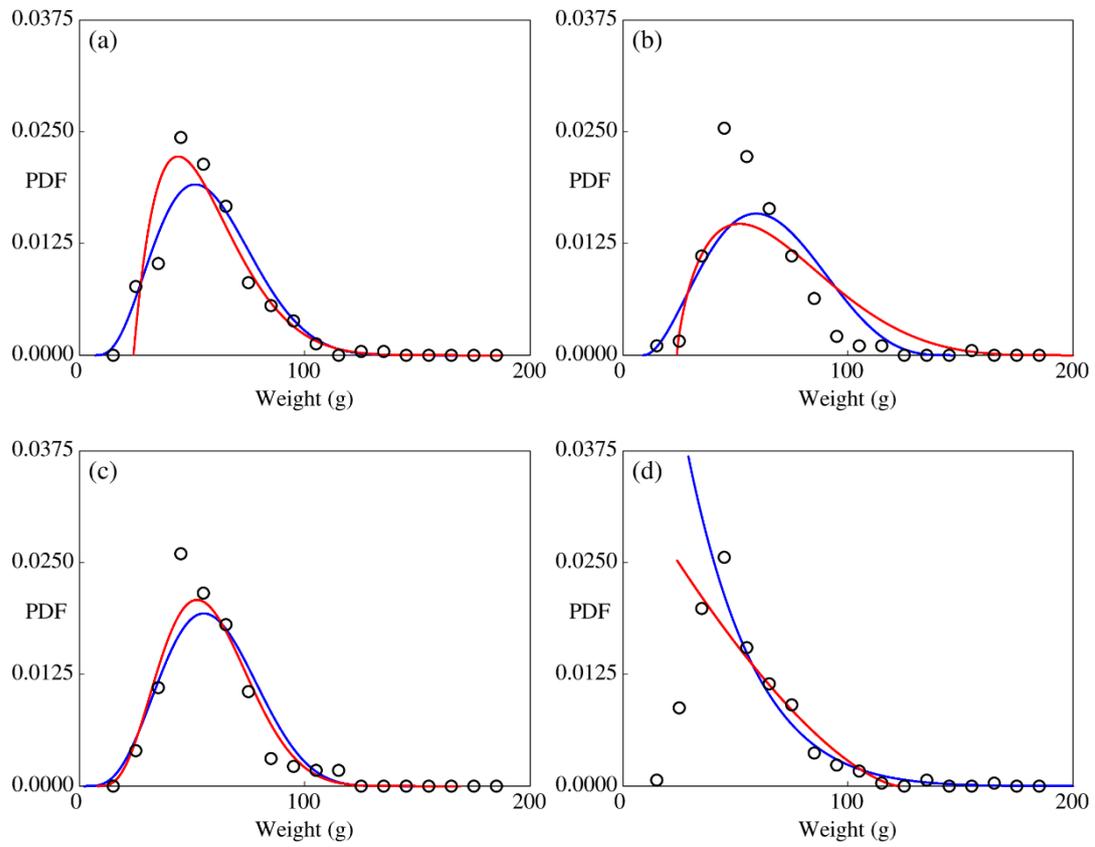

**Figure 6.7.** Empirical body weight distribution (circles) and theoretical results with the fitted uncertain logistic model (curves) at the Toami competition in each year: the model with $W_0 = 10$ (g) (blue) and $W_0$ in **Table 6.1** (red).

**6.2.4. Review of *P. altivelis* growth**

This section reviews the growth of *P. altivelis* from spring to autumn in other rivers in Japan through a literature survey. Many studies have used the standard length of the fish, while what we collected in the Hii River was body weight. This is mainly because body weight is more related to resource biomass and partly to the technical assumptions of our surveys. Therefore, we reviewed the literature in which the body weights of caught fish were recorded. The results are summarized in **Table 6.5**, which are not exhaustive but provide insights into the body weight data of the fish at our study site.

We give several remarks below. Mori et al. (2008) investigated density effects of the fish in a river, concluding that there is a negative relationship between the average body weight of the fish and their population density. According to the growth data of Tezuka and Takeda (2003) for different years, the growth dynamics of the fish are significantly different among years with the maximum difference about 40 (g). Nakagawa et al. (2015) showed that the standard deviation of the body weight is at the same order with the mean, both being $O(10^1)$ (g). They also reported that the difference among sites is also significant. The data of Katun et al. (2023) show that the mean body weight of the fish is around 30 (g) from August to October, which is about half of ours. These findings imply that our data and model identification results that the models are qualitatively different, like the beta distribution profiles, among different years is reasonable, and further suggest that accounting for individual differences is essential for analyzing biological growth of the fish.

**Table 6.5.** Body weight data of *P. altivelis* in Japan.

| Water body | Season | Body weight | Source and Remark |
|---|---|---|---|
| Umi River, Niigata Pref. | Around August 1 | 9.5 (g) to 31.2 (g) | Mori et al. (2008) |
| Naka River, Tochigi Pref. | March 21 | 3.83±0.43 (g) | Tezuka and Nakamura (1994) Released fish |
| Naka River, Tochigi Pref. | June to October | 40 (g) to 80 (g) depending on year | Tezuka and Takeda (2003) |
| Naka River, Tochigi Pref. | Late June to early July | 51.0 (g) to 80.9 (g) | Takagi et al. (2017) Recaptured |
| Yahagi River, Aichi Pref. | July to the beginning of September | From 13.5 (g) to 104.6 (g) Ave. 52.6 (g) | Uchida (2002) |
| Sho River, Toyama Pref. | June to July | 10.2 (g) to 79.1 (g) | Tago (2003) |
| Sho River, Toyama Pref. | After September in several years | Around 5 (g) to 25 (g) | Iguchi et al. (2011) |
| Toyama Bay | March to April | 1.1 (g) to 7.2 (g) | Tago (2002) |
| Kamo River, Kyoto Pref. | Around Autumnn | Mid reaches: 66±22 (g) Lower reaches: 53±17 (g) | Nakagawa et al. (2015) |
| Takahashi River, Okayama Pref. | April 13 | Natural: 2.87±0.74 (g) Released: 5.97±0.49 (g) | Kobori and Abe (2019) |
| Hirose River, Miyagi Pref. | May to June | 1.85 (g) to 45.30 (g) | Watanabe et al. (2021) |
| Gono River, Hiroshima Pref. | Throughout a year | Around 30 (g) during August to October | Katun et al. (2023) |

## 6.3 Optimal control modeling

### 6.3.1 System dynamics

Typically, a harvesting problem of a fishery resource is based on its biomass or population in some habitat, which are difficult to continuously observe in the real world. For example, one may design a harvesting policy for some fish in a river based on the total number of individual fishers in it, which is difficult to observe and estimate even if not possible. Moreover, population dynamics of fishes in inland water bodies have been suggested to vary dynamically, such as about twenty times within successive ten years (Mori et al., 2008; Yoshioka et al., 2023). In many rivers including the Hii River that inland fishery cooperatives artificially release fishes from farms in each year, and their population is known. Moreover, it can be easier to track cumulated caught fishers than counting fish population in a water body. Based on these considerations, we present a simple but possibly more realistic model in this view for designing harvesting policy of the fish *P. altivelis* during a fishing season. The model presented below is a simplified version of that in Yoshioka (2024b). An advantage of the present model over the earlier model is that its temporal discretization can be conducted without resorting nonlinear numerical solvers.

We consider the following ODE:

$$\underbrace{\mathrm{d}N_t}_{\text{Population change}} = \underbrace{-q_t \mathrm{d}t}_{\text{Decay by harvesting}} \quad \text{for } t > 0 \quad \text{subject to} \quad N_0 > 0 \quad (6.8)$$

with a nonnegative control $q = (q_t)_{t \geq 0}$ such that the natural state constraint $N_t \geq 0$ ($t \geq 0$) is satisfied. The initial condition $N_0$ is the fish that has been known to exist at the time zero. For example, it can be set as the total number of released fishes at time zero. The population should be a nonnegative integer in the real world, but we assume it to be a continuous variable assuming that it is significantly larger than $O(10^0)$, e.g., $O(10^{4\text{-}5})$.

Possibly there are natural fish population in the same water body as well. In such as case, this $N_0$ serves as a lower bound of the fish population that can be potentially harvested. Of course, there will be population decay as the time elapses, which will be considered in the objective function but not the dynamics (6.8) as a discount factor. An important point in the proposed population dynamics is that in practice one can count the unit-time harvested population, which is $q_t$ at each time, and hence $N_t$. Our modeling strategy is therefore designing a harvesting policy not based on the (unknown) existing population in the habitat but rather on the total harvested one.

### 6.3.2 Control problem

The objective to be maximized is set with a fixed terminal time $T > 0$ as follows:

$$J(t, n, q) = \int_t^T \exp(-\delta(s-t)) \left( \underbrace{\frac{1}{\alpha}(\omega(s, N_s) q_s)^\alpha}_{\text{Harvesting utility}} - \underbrace{h q_s}_{\text{Harvesting cost}} \right) \mathrm{d}s + \underbrace{\exp(-\delta(T-t)) S(N_T)}_{\text{Sustainability concern}} \quad (6.9)$$

with the entropic lower bound of the average body weight under model uncertainty

$$\omega(t,n) = -\frac{1}{\eta(n)} \ln \left\{ \int_{\Omega_W} \exp(-\eta(n) W(t, W_{max})) P(dW_{max}) \right\}, \tag{6.10}$$

where

$$W(t, W_{max}) = \frac{W_{max}}{1 + \left(\frac{W_{max}}{W_0} - 1\right) \exp(-rt)} \tag{6.11}$$

and a state-dependent uncertainty-aversion parameter $\eta(n) > 0$. In (6.9), $\delta \geq 0$ is the discount rate that can be effectively considered as a risk factor in the proposed model such that a larger value implies an earlier termination of harvesting: the harvesting period effectively terminates at a random time following the exponential distribution $\delta \exp(-\delta \tau)$ ($\tau > 0$). In the objective function (6.9), the first term is the harvesting cost utility with the shape parameter $\alpha \in (0,1)$ meaning that the utility is concave with respect to the harvested biomass, the second term is the harvesting cost with a proportional constant $h > 0$ meaning that the harvesting cost is proportional to the harvested population, and the third term represents sustainability concern with a non-negative increasing function $S : [0,+\infty) \to [0,+\infty)$ bounded from above by a constant $\bar{S} > 0$. The increasing property of $S$ means that the sustainability concern is better resolved if the population is less exploited during the period $[0,T]$. We assume in the sequel that $S(0) = 0$, implying that nothing is obtained at the terminal time if there is no population at that time.

The decision-maker to optimize the objective function (6.9) is assumed to be an aggregation of anglers; therefore, it is more reasonable to formulate the control problem in a mean field where a larger number of anglers interact with each other. This is theoretically possible; however, the tractability of our modeling framework will be lost and will be studied elsewhere. The use of the lower bound (6.10) enables us to account for misspecification in biological growth, which is modeled in our context through a distortion of the probability density $P(dW_{max})$. The entropic lower bound (6.10) is equivalently written as

$$\omega(t,n) = \inf_{\phi} \left\{ \underbrace{\int_{\Omega_W} \phi(t, W_{max}) W(t, W_{max}) P(dW_{max})}_{\text{Distorted average weight}} - \underbrace{\frac{1}{\eta(n)}}_{\text{Weighting constant}} \underbrace{\int_{\Omega_W} (\phi(t, W_{max}) \ln \phi(t, W_{max}) - \phi(t, W_{max}) + 1) P(dW_{max})}_{\text{Relative entropy}} \right\}, \tag{6.12}$$

where the infimum is considered with respect to measurable functions $\phi : \Omega_W \to [0,+\infty)$ such that $\int_{\Omega_W} \phi(t, W_{max}) P(dW_{max}) = 1$. The minimizer $\phi = \phi^*$ in (6.12) is determined by an optimization subject to the integral constraint as follows:

$$\phi^*(t,n) = \frac{\exp\{-\eta(n) W(t, W_{max})\}}{\int_{\Omega_W} \exp\{-\eta(n) W(t, W_{max})\} P(dW_{max})}. \tag{6.13}$$

The worst-case distorted body weight distribution then becomes $\phi^*(t, W_{max}) P(dW_{max})$. The second term on the right-hand side of (6.12) is a version of the divergence accounting for possible misspecification in such a way that there is some misspecification if $\phi(t, W_{\max})$ is not identical to the constant function 1. The weighting constant $\eta(n)$, the state- and hence population-dependent uncertainty-aversion parameter, modulates the uncertainty in the biological growth model such that a larger value of $\eta(n)$ implies a larger uncertainty and vice versa. Since the objective here is to maximize the harvesting utility, it is safer to assume that uncertainties act to underestimate the total harvested biomass, which is the reason that we consider "inf" in the formulation (6.12).

Finally, the value function $\Phi: [0,T] \times [0,+\infty) \to \mathbb{R}$, the optimized objective function, is given by

$$\Phi(t,n) = \sup_q J(t,n,q). \tag{6.14}$$

Here and in the sequel, control $q$ is assumed to be a measurable function such that the dynamics (6.8) admits a unique continuous solution $N = (N_t)_{0 \le t \le T}$. The goal of the control problem is to find a maximizing $q = q^*$ of (6.14) as well as $\Phi$ itself. The dynamic programming approach discussed in **Chapter 2** plays a vital role.

Finally, we have the boundary condition $\Phi(t,0) = 0$ ($0 \le t \le T$) due to $S(0) = 0$ and $J(t,0,q) = 0$ because only the null control $q \equiv 0$ is admissible considering the dynamics (6.8). In the sequel, we choose $\alpha = 1/2$, with which we can present an analytically tractable HJB equation as well as its numerical schemes. The cases $\alpha \neq 1/2$ are more complicated and the proposed numerical method does not apply.

### 6.3.3 HJB equation

The HJB equation associated with the control problem is formally given by

$$\begin{aligned}-\frac{\partial \Phi(t,n)}{\partial t} &= \sup_{q \ge 0} \left\{ \delta \Phi(t,n) - q \frac{\partial \Phi(t,n)}{\partial n} + 2\sqrt{\omega(t,n)q} - hq \right\} \\ &= \mathbb{H}\left(t, n, \Phi(t,n), \frac{\partial \Phi(t,n)}{\partial n}\right)\end{aligned}, \quad 0 \le t < T \text{ and } n > 0 \tag{6.15}$$

with the Hamiltonian $\mathbb{H}: [0,T] \times [0,+\infty) \times \mathbb{R} \times \mathbb{R} \to \mathbb{R}$ of the form

$$\mathbb{H}(t,n,u,p) = -\delta u + \frac{\omega(t,n)}{h+p}. \tag{6.16}$$

The HJB equation (6.15) is equipped with the terminal condition

$$\Phi(T,n) = S(n), \quad n \ge 0 \tag{6.17}$$

and the boundary condition

$$\Phi(t,0) = 0, \quad 0 \le t < T. \tag{6.18}$$

The terminal and boundary conditions are compatible because of the assumption $S(0) = 0$. The Hamiltonian $\mathbb{H}$ is not definable when $h + p = 0$, but is found not to occur for the HJB equation (6.15). (see **Appendix**).

The quantity (6.10) is an entropic lower bound of the average body weight under uncertainty with a state-dependent uncertainty-aversion parameter $\eta(n) > 0$. The Hamiltonian $\mathbb{H}$ is decreasing with respect to $p \geq 0$, which is crucial for construction stable numerical schemes for discretization of the HJB equation (6.15). The (guessed) optimal control associated with the HJB equation (6.15) is given by

$$q_t^* = \frac{\omega(t, N_t)}{\left(h + \frac{\partial \Phi(t, N_t)}{\partial n}\right)^2} \quad \text{if } N_t > 0 \text{ and } q_t^* = 0 \text{ otherwise} \tag{6.19}$$

if we always have $\frac{\partial \Phi}{\partial n} \geq 0$ in some sense, which is justified in **Appendix**. In this view, we can concurrently find the value function and the optimal by solving the HJB equation (6.15).

The output from our HJB equation is not only the value function and optimal control, but also the worst-case distortion depending on the harvesting trajectory considering (6.13):

$$\phi^*(t, N_t, W_{\max}) = \frac{\exp\{-\eta(N_t) W(t, W_{\max})\}}{\int_{\Omega_W} \exp\{-\eta(N_t) W(t, W_{\max})\} P(\mathrm{d}W_{\max})}. \tag{6.20}$$

Notice that the population $N$ here depends on $q^*$ and hence on $\Phi$ through the relationship (6.19).

### 6.3.4 Numerical discretization

We examine three numerical schemes to discretize the HJB equation (6.15) because we have not been aware of its analytical solutions. The equation must be solved backward in time because it is a terminal value problem. Therefore, solutions are constructed from the future to the past.

We discretize the space-time domain $[0, T] \times [0, N_{\max}]$, where $N_{\max} > 0$ is some prescribed constant as follows. We define vertices $P_{i,j} : (t_i, n_j)$ ($i = 0, 1, 2, .., I_t$ and $j = 0, 1, 2, .., I_n$) with some $I_t, I_n \in \mathbb{N}$, the computational resolution as the spacings between each successive vertices in the directions of $t$ and $n$ given by $\Delta t = T / I_t$ and $\Delta n = N_{\max} / I_n$, respectively. We have also set $t_i = i \Delta t$ and $n_j = j \Delta n$. The discretized value function $\Phi$ at $P_{i,j}$ is expressed as $\Phi_{i,j}$, and we call its collection at all vertices $P_{i,j}$ as a numerical solution. The terminal condition is $\Phi_{I_t, j} = S(n_j)$ ($j = 0, 1, 2, .., I_n$).

The numerical schemes considered in this chapter are presented as follows. All the schemes share the common discretization of the boundary condition for $i = 0, 1, 2, .., I_t$ as $\Phi_{i,0} = 0$. The explanation below therefore considers $i = 0, 1, 2, .., I_t - 1$ and $j = 1, 2, 3, .., I_n$.

**Explicit scheme**

The simplest scheme will be an explicit scheme that evaluates the right-hand side of the HJB equation based only on the available values at the previous (future) time step:

$$-\frac{\Phi_{i,j} - \Phi_{i-1,j}}{\Delta t} = \mathbb{H}\left(t_{i-1}, n_j, \Phi_{i,j}, \frac{\Phi_{i,j} - \Phi_{i,j-1}}{\Delta n}\right) = -\delta \Phi_{i,j} + \frac{\omega(t_{i-1}, n_j)}{h + \frac{\Phi_{i,j} - \Phi_{i,j-1}}{\Delta n}} \quad (6.21)$$

or equivalently

$$\Phi_{i-1,j} = \Phi_{i,j} + \Delta t \mathbb{H}\left(t_{i-1}, n_j, \Phi_{i,j}, \frac{\Phi_{i,j} - \Phi_{i,j-1}}{\Delta n}\right). \quad (6.22)$$

The scheme is stable (i.e., numerical solutions do not artificially blow up) only if the time increment $\Delta t$ is chosen sufficiently small. An advantage of the scheme is that the solution procedure is "explicit" in accordance with its name; it does not use any matrix inversion.

**Semi-implicit scheme**

The explicit scheme can be modified without losing the "explicit" nature. We present two such schemes in the sequel. The first of them is the semi-implicit scheme below where only the linear term is implicitly discretized:

$$-\frac{\Phi_{i,j} - \Phi_{i-1,j}}{\Delta t} = -\delta \Phi_{i-1,j} + \frac{\omega(t_{i-1}, n_j)}{h + \frac{\Phi_{i,j} - \Phi_{i,j-1}}{\Delta n}} \quad (6.23)$$

or equivalently

$$\Phi_{i-1,j} = (1 + \delta \Delta t)^{-1}\left[\Phi_{i,j} + \Delta t \frac{\omega(t_{i-1}, n_j)}{h + \frac{\Phi_{i,j} - \Phi_{i,j-1}}{\Delta n}}\right] \quad (6.24)$$

that can be analytically solved from $j = 0$ to $j = I_n$ at each time step in a cascading manner. Therefore, by specifying the direction of updating numerical solutions, the "explicit" nature is not lost. The computational complexity is therefore at the same level as the explicit scheme. However, we can also discretize the nonlinear time implicitly as shown below. The semi-implicit scheme is expected to be more stable than the explicit one because of partly employing an implicit discretization.

**Implicit scheme**

The third scheme is the implicit scheme where both the linear and nonlinear terms in the right-hand side are discretized implicitly:

$$-\frac{\Phi_{i,j} - \Phi_{i-1,j}}{\Delta t} = \mathbb{H}\left(t_{i-1}, n_j, \Phi_{i-1,j}, \frac{\Phi_{i-1,j} - \Phi_{i-1,j-1}}{\Delta n}\right) = -\delta \Phi_{i-1,j} + \frac{\omega(t_{i-1}, n_j)}{h + \frac{\Phi_{i-1,j} - \Phi_{i-1,j-1}}{\Delta n}} \quad (6.25)$$

or equivalently

$$C_A \left(\Phi_{i-1,j}\right)^2 + C_B \Phi_{i-1,j} - C_C = 0 \tag{6.26}$$

that can be solved as (we take the larger root in the quadratic equation)

$$\Phi_{i-1,j} = \frac{-C_B + \sqrt{C_B^2 + 4C_A C_C}}{2C_A}, \tag{6.27}$$

where the coefficients $C_A, C_B, C_C$ are given by

$$C_A = 1 + \delta t \geq 1, \quad C_B = C_A \left(h\Delta n - \Phi_{i-1,j-1}\right) - \Phi_{i,j}, \quad C_C = \omega(t_{i-1}, n_j)\Delta t \Delta n + \left(h\Delta n - \Phi_{i-1,j-1}\right)\Phi_{i,j}. \tag{6.28}$$

Here, we have

$$\begin{aligned} C_B^2 + 4C_A C_C &= \left\{C_A\left(h\Delta n - \Phi_{i-1,j-1}\right) - \Phi_{i,j}\right\}^2 + 4\omega(t_{i-1}, n_j)\Delta t \Delta n + 4C_A\left(h\Delta n - \Phi_{i-1,j-1}\right)\Phi_{i,j} \\ &= \left\{C_A\left(h\Delta n - \Phi_{i-1,j-1}\right) + \Phi_{i,j}\right\}^2 + 4\omega(t_{i-1}, n_j)\Delta t \Delta n \\ &\geq 0 \end{aligned} \tag{6.29}$$

so that (6.27) is well-defined. We point out that the other root of the quadratic equation (6.26) is improper as a numerical solution because it is not monotone (**Appendix**), which is a desirable property to be equipped with numerical schemes for discretizing HJB and related equations.

Like the semi-implicit scheme, the implicit scheme can be analytically solved from $j = 0$ to $j = I_n$ in a cascading manner at each time step. The scheme is seemingly the most complex among the three schemes, while it is essentially a solution of a quadratic equation whose solutions are necessarily real variables. Because of the formula (6.27), its computational implementation would not face technical difficulties. Therefore, we consider the three schemes to be comparable in terms of computational complexity. We show that the implicit scheme is the most stable among the three schemes in the sense that it is unconditionally table irrespective of $\Delta t$ and $\Delta n$.

In the **Appendix**, we theoretically analyze the HJB equation and its discretization schemes. The machinery here is to check the monotonicity, stability, and consistency with which the convergence of numerical solutions to a scheme to a viscosity solution (a proper weak solution that is not always partially differentiable everywhere) to our HJB equation. Moreover, if the HJB equation admits a unique continuous viscosity solution, the numerical solutions converge under mild conditions.

**6.4 Application study**

We compared the three numerical schemes for the discretization of our HJB equation with the uncertain logistic growth models identified in this chapter. Unless otherwise specified, we used the model for the year 2023 for the uncertain logistic growth model with the initial body weight presented in **Table 6.1**. The other model parameters are chosen for demonstrative purposes, so that one can visually understand the properties of the solution to our HJB equation and numerical schemes. The time interval $[0,T]$ is the 120 days starting from July 1 at which harvesting of *P. altivelis* in the Hii River starts. The computational resolution is chosen as $I_t = 24,000$ and $I_n = 500$ with the normalization $N_{\max} = 1$ without loss of any theoretical generalities. For the other parameter values, we set the discount rate $\delta = 0.04$ (1/day) and the harvesting cost $h = 100$, and $\mu = 0.1$, with which we obtain nontrivial computational results as demonstrated below.

We firstly in vestige the case when there is no sustainability concern $S(n) \equiv 0$ as shown in **Figure 6.8** (value functions and corresponding controlled trajectories) and **Figure 6.9** (optimal harvesting policies and corresponding controlled trajectories). The trajectories were chosen according to their terminal values and simulated using backward tracking. The three numerical schemes provide comparable computational results that are difficult to distinguish visually. When there is no sustainability concern, the controlled population $N$ decreases without inflection points because harvesting the fish, even their runout, incurs nothing at the end of the harvesting period. Both the computed value function and optimal control were smooth. In contrast, **Figure 6.10** (value functions and corresponding controlled trajectories) and **Figure 6.11** (optimal harvesting policies and corresponding controlled trajectories) for the following discontinuous sustainability concern yield critically different results with each other: $S(n) = 50$ if $n \geq 1/2$ and $S(n) = 0$ otherwise. In the discontinuous case, the continuity assumption of the terminal condition is violated, but the computation is successfully conducted. As shown in **Figures 6.10 and 6.11**, controlled trajectories starting from a sufficiently large initial condition do not have inflection points as in the previous case, and gain a positive terminal profit, while those with moderately large initial conditions do. More specifically, some trajectories have inflection points following the line $n = 1/2$ along which the sustainability concern is discontinuous. These trajectories first decrease the population, then almost stop harvesting, and then again start harvesting. The no-harvesting period partly contributes to the sustainability of fishery resources. It should be noted that trajectories with a sufficiently small initial condition never stop harvesting because it is more profitable for them to continue harvesting than to do nothing, considering resource sustainability. Again, the three schemes perform comparably.

We also investigated the differences among the numerical solutions to the three schemes. **Figure 6.12** shows the difference of value functions between the explicit to semi-implicit (e.g., value of the explicit scheme minus that of the semi-implicit scheme), explicit to implicit, and semi-implicit to implicit schemes when there is no sustainability concern. Similarly, **Figure 6.13** shows the results for discontinuous sustainability concerns examined above. **Figure 6.12** shows the case with no sustainability concern, where the explicit scheme is more optimistic among the three schemes because it gives the largest computed value function, followed by the semi-implicit and implicit ones in this order. In this case, the implicit nature stabilized the solution the most; therefore, it is considered to induce the largest numerical diffusion. Such a clear trend is not observed for the case with discontinuous sustainability concern in **Figure 6.13**. Nevertheless, as visually suggested in **Figures 6.8 through 6.11**, the computational performance of the three schemes is almost the same, and any of them can be utilized under this computational resolution. We also examine the same analysis with the lower resolution $I_t = 4,800$ and $I_n = 100$ for the case with no sustainability concern as shown in **Figure 6.14**. The computational results in this figure suggest that the difference among the schemes becomes three to five times larger for the coarser resolution, but still small because the value function is at the order of $O(10^0)$.

Finally, we conducted a sensitivity analysis using the semi-implicit scheme for the case with a discontinuous sustainability concern. **Figure 6.15** shows the computed value functions and controlled trajectories: (a) nominal case (the parameters specified above), (b) larger discount case with $\delta = 0.08$, (c) larger state-dependent model uncertainty 1 ($\eta(n) = (1-9n/10)/10$), and (d) larger state-dependent model uncertainty 2 ($\eta(n) = (1+n)/10$). Increasing the discount rate yields a more aggressive harvesting policy, owing to the possible earlier random termination of harvesting. Increasing model uncertainty through the growth model yields less aggressive harvesting policies. To deeper investigate influences of model uncertainties, the distorted probability distributions of the maximum body weight, which is $\phi_t^* p(w)$, are investigated along the trajectory terminating at $1/2$. **Figure 6.16** shows that the computational results are identical between the nominal and large-discount cases, whereas the difference between the nominal and uncertain cases is visible. Assuming a larger model uncertainty yields a probability distribution of the maximum body weight that is more concentrated near the left boundary. It is also interesting to observe that in each case, the distribution is more concentrated on the left boundary as time elapses, suggesting that the harvested biomass is more underestimated at times closer to the terminal time in the proposed model, although the body weight itself will become larger on average.

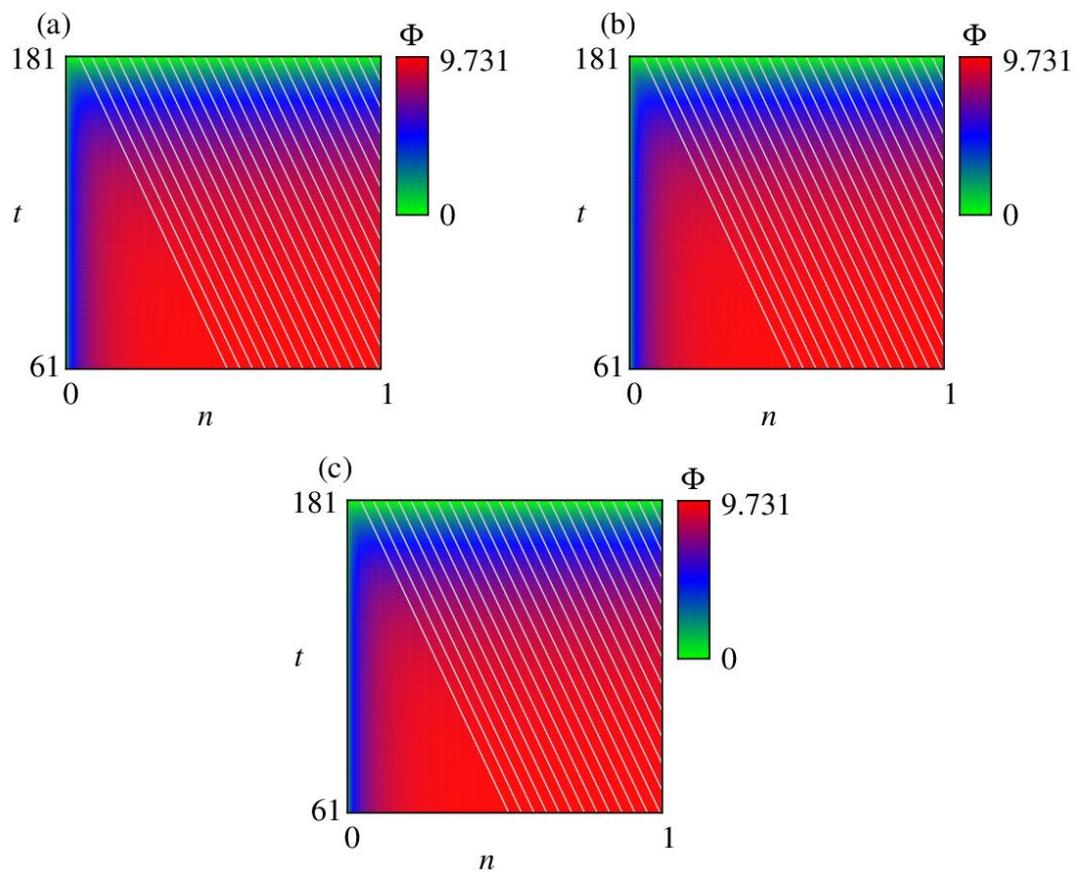

**Figure 6.8.** Computed value functions (heat map) and resulting controlled trajectories (curves) for no sustainable concern: (a) Explicit scheme, (b) Semi-implicit scheme, and (c) Implicit scheme.

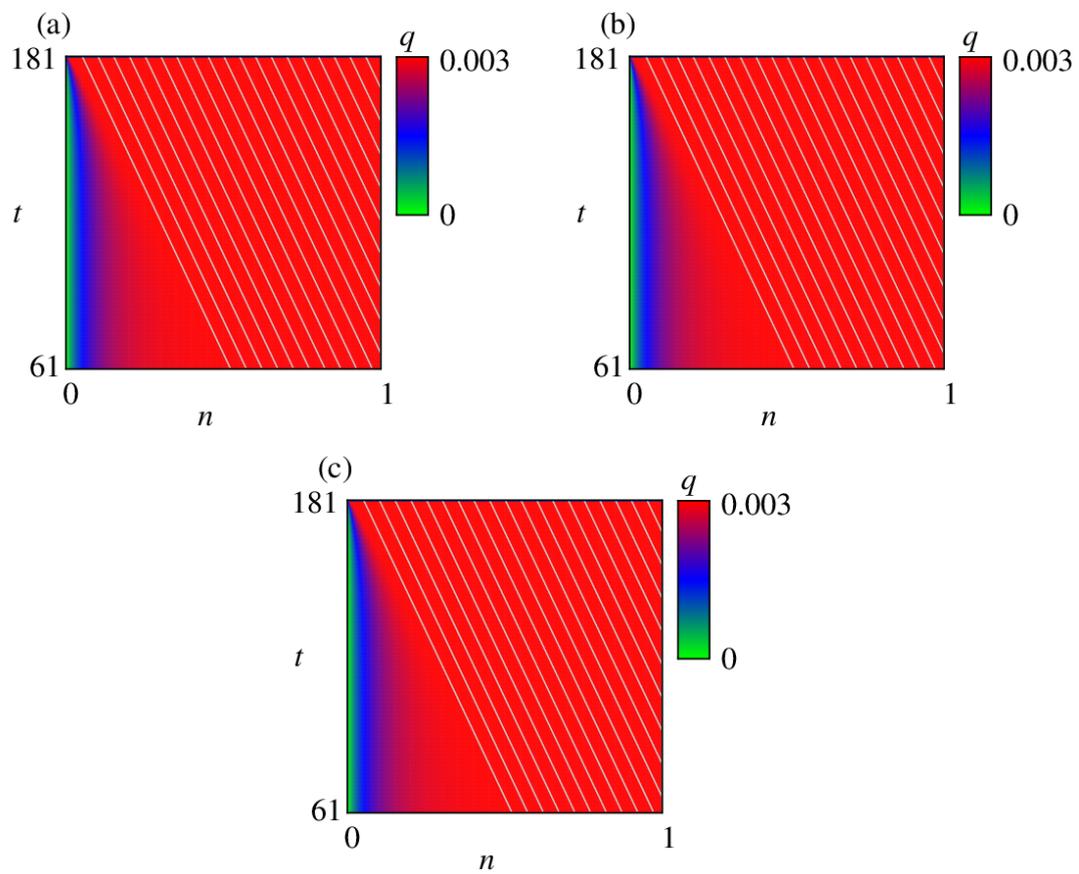

**Figure 6.9.** Computed optimal harvesting policies (heat map) and resulting controlled trajectories (curves) for no sustainable concern: (a) Explicit scheme, (b) Semi-implicit scheme, and (c) Implicit scheme.

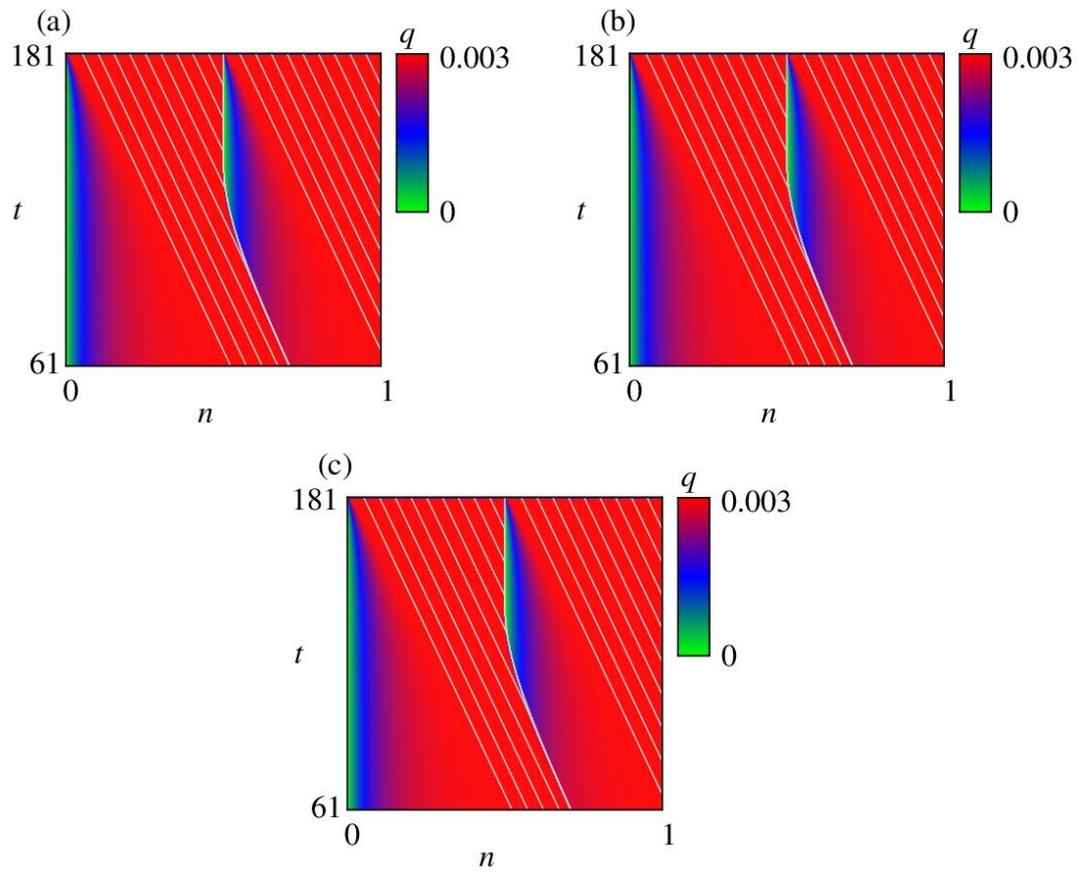

**Figure 6.10.** Computed value functions (heat map) and resulting controlled trajectories (curves) for the discontinuous sustainable concern: (a) Explicit scheme, (b) Semi-implicit scheme, and (c) Implicit scheme.

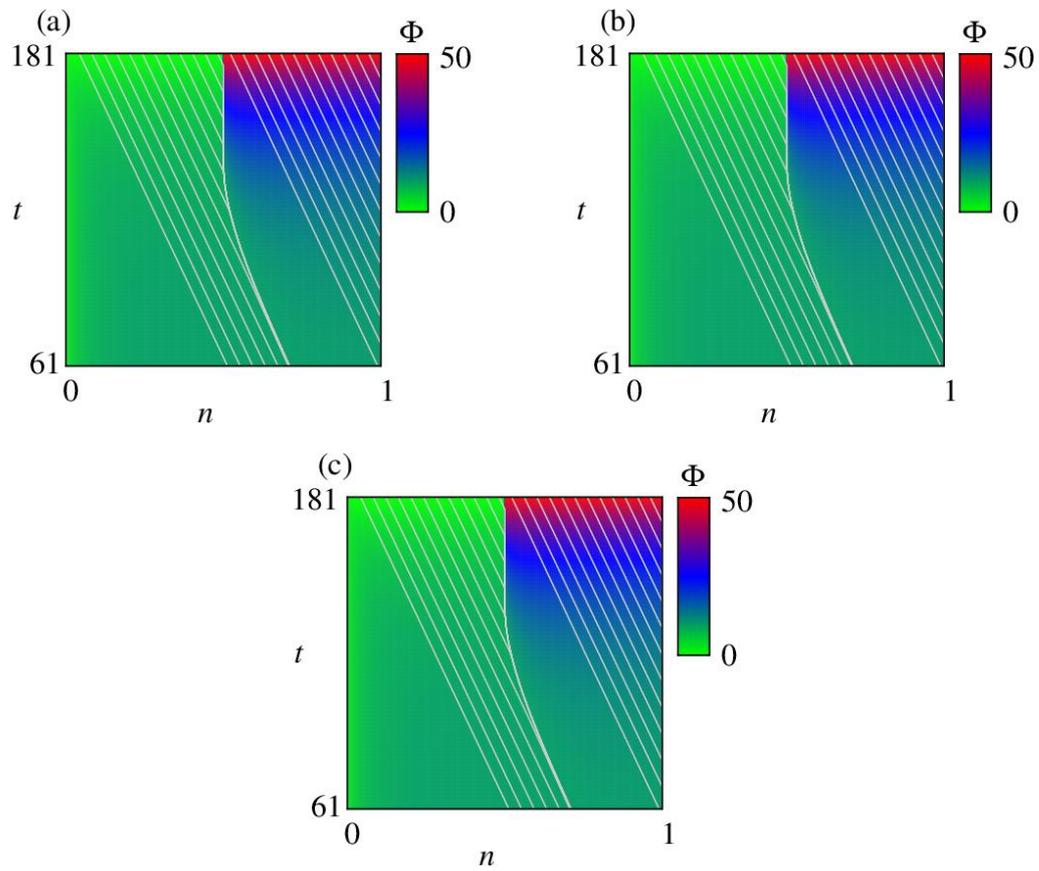

**Figure 6.11.** Computed optimal harvesting policies (heat map) and resulting controlled trajectories (curves) for discontinuous sustainable concern: (a) Explicit scheme, (b) Semi-implicit scheme, and (c) Implicit scheme.

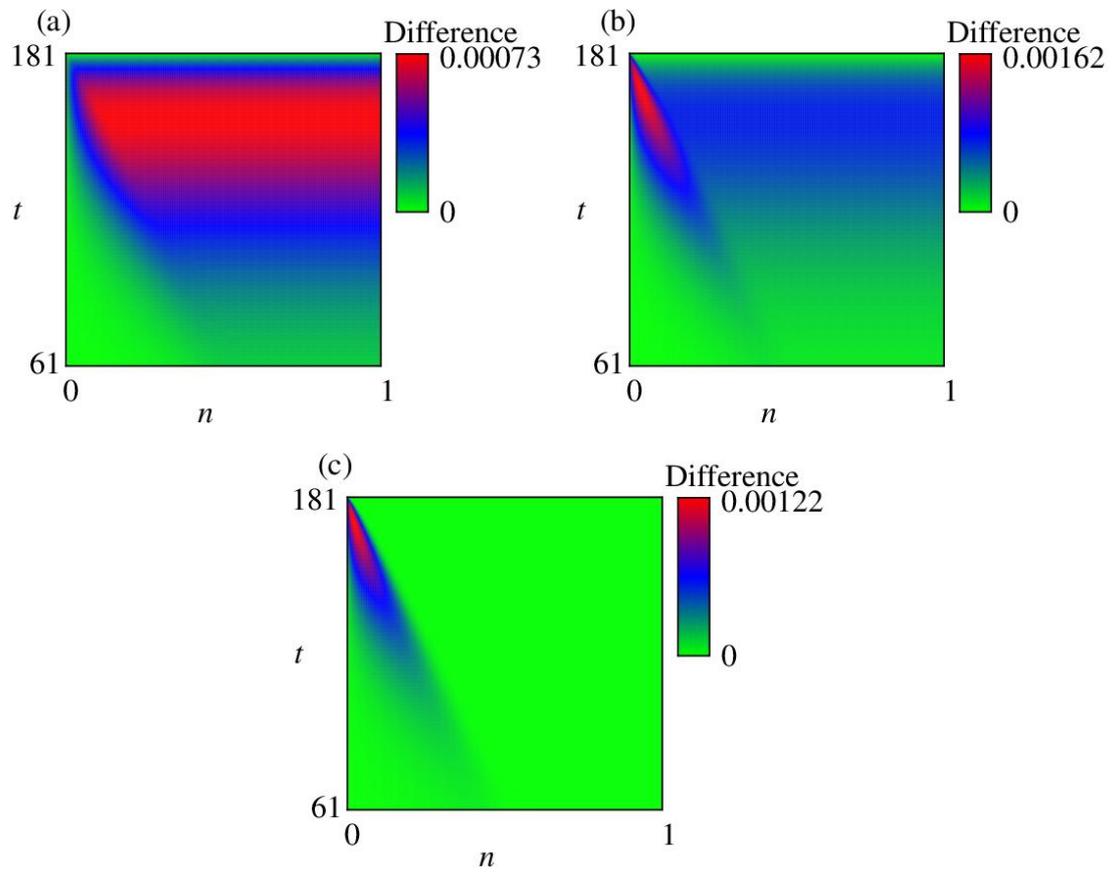

**Figure 6.12.** Difference in value functions between (a) explicit to semi-implicit schemes, (b) explicit to implicit schemes, and (c) semi-implicit to implicit schemes when there is no sustainability concern.

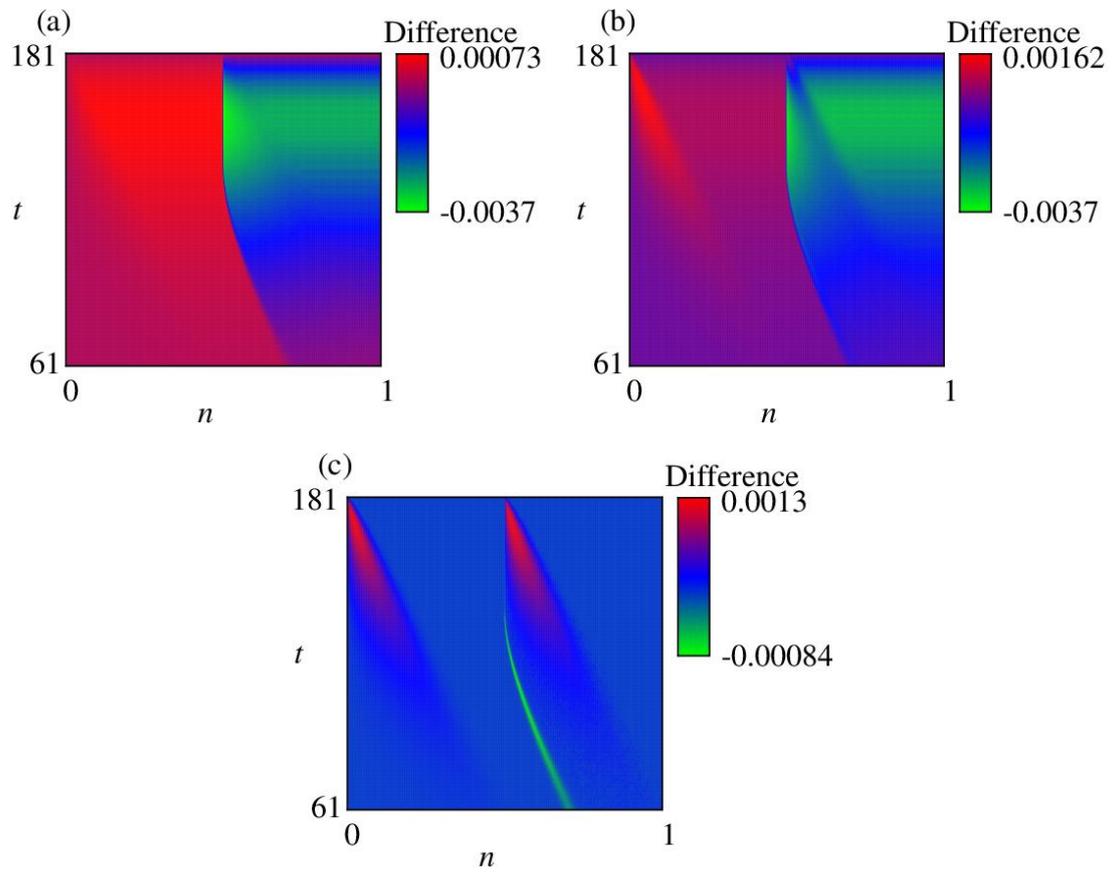

**Figure 6.13.** Difference in value functions between (a) explicit to semi-implicit schemes, (b) explicit to implicit schemes, and (c) semi-implicit to implicit schemes with a discontinuous sustainability concern.

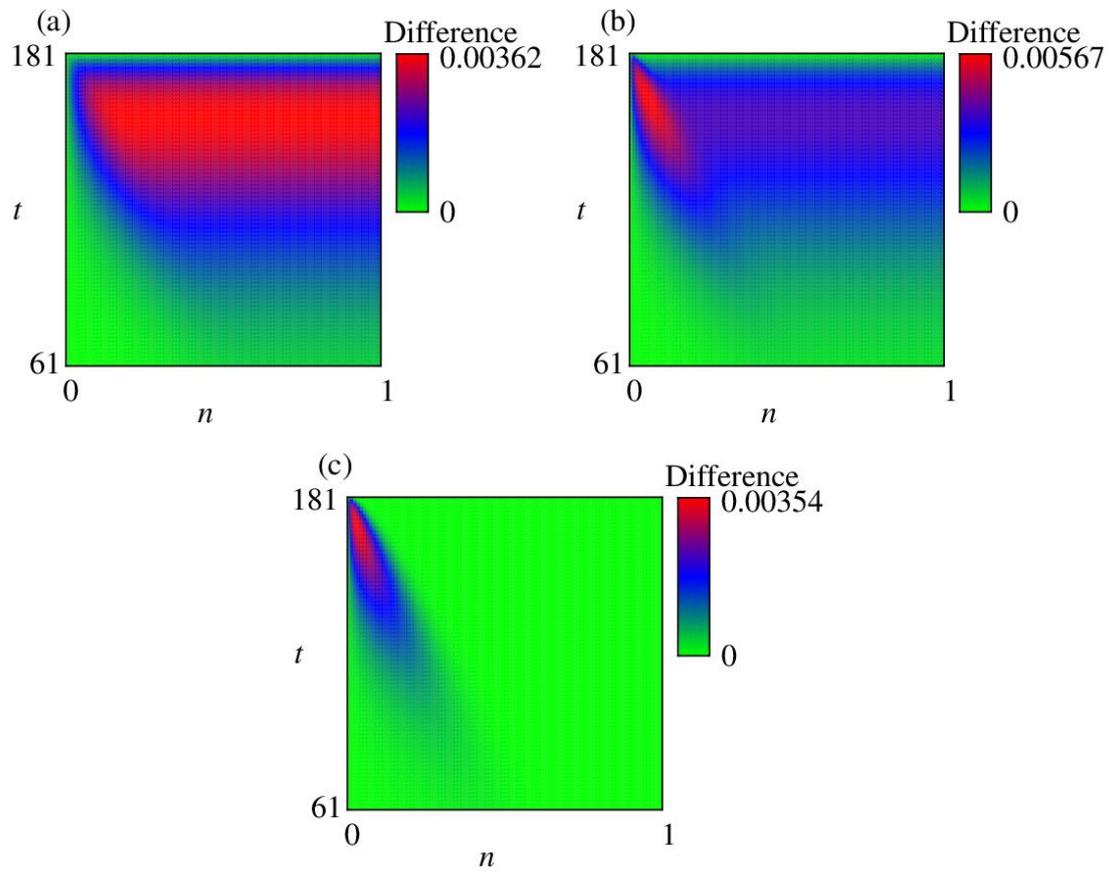

**Figure 6.14.** Same as **Figure 6.12** but with a coarser resolution.

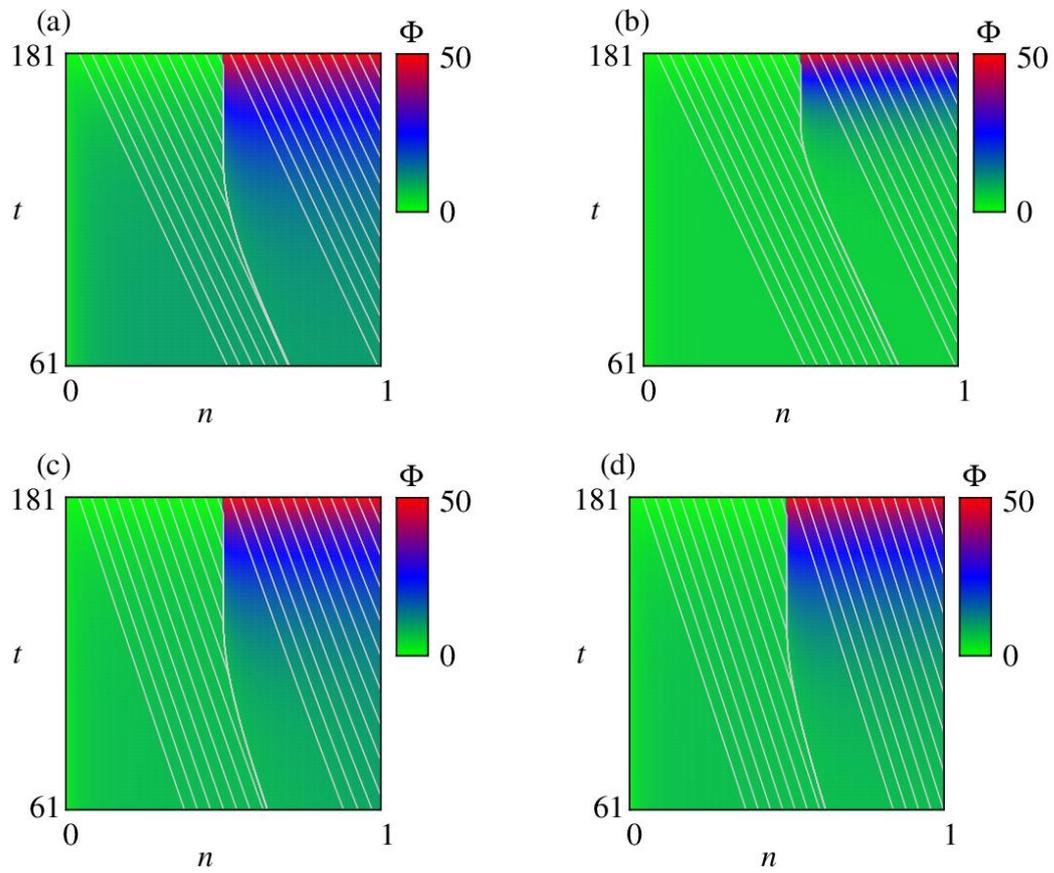

**Figure 6.15.** Sensitivity analysis of the value functions and controlled trajectories: (a) nominal case, (b) larger discount case, (c) larger state-dependent model uncertainty 1, and (d) larger state-dependent model uncertainty 2.

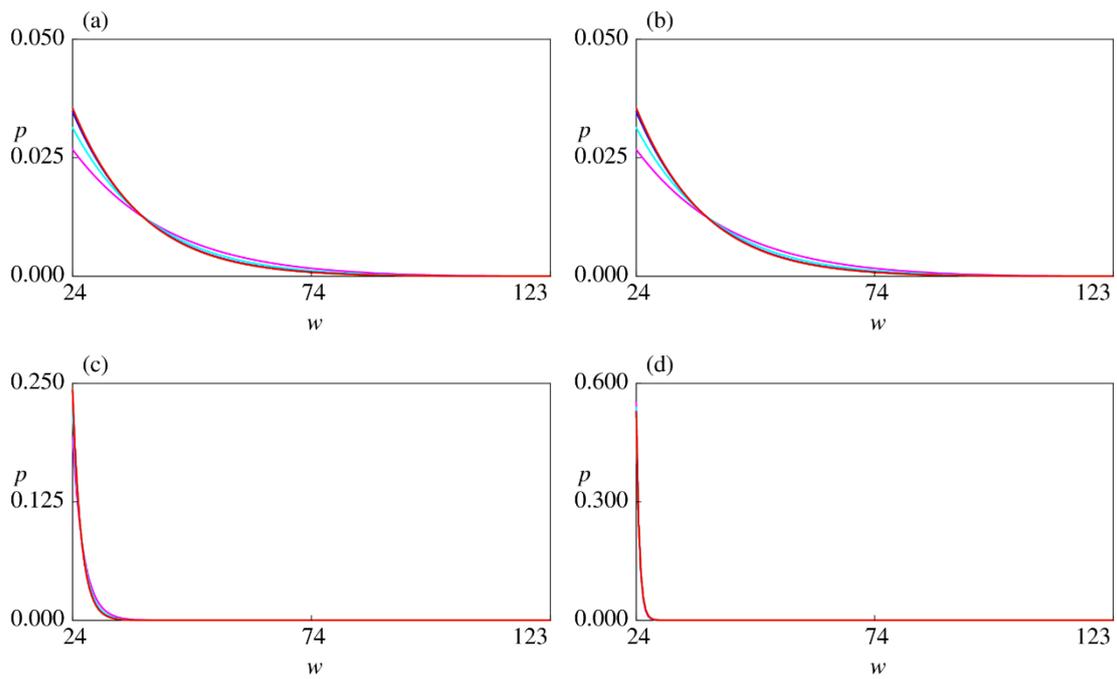

**Figure 6.16.** Sensitivity analysis of the distorted probability distribution of the maximum body weight along the trajectory terminating at $1/2$ : (a) nominal case, (b) larger discount case, (c) larger state-dependent model uncertainty 1, and (d) larger state-dependent model uncertainty 2. The colors represent the distribution at time 61 day (magenta), 91 (day) (light blue), 121 (day) (blue), 151 day (green), and 180 (day) (red).

**6.5 Conclusion**

This chapter summarized the biological growth data for *P. altivelis* collected from the Hii River, Japan. The uncertain logistic-growth model was fitted against the data, and yearly differences were discussed. An optimal control model based on an uncertain logistic model and a simple population dynamics model was then proposed to analyze sustainable inland fisheries at the study site. Finite difference methods were applied to the discretization of the control problem, successfully generating stable numerical solutions. The computational results suggest that the three numerical schemes introduced in this chapter perform comparably, with slight differences in the theoretical estimates, as discussed in the **Appendix**.

We did not consider cases where a fish was captured multiple times because *P. altivelis* is so weak that it cannot survive once an angler catches it. However, some larger and physiologically stronger fish species may be caught multiple times during their life history, the so-called catch and release angling. In such cases, we need to reconsider both angler and fish population dynamics, such that multiple catch events of each individual fish occur, by accounting for size-selective harvesting (Carosi et al., 2022) and the risks of fishers being caught multiple times (Thorstad et al., 2019). We believe that our mathematical and computational approaches will theoretically carry over to these advanced problems using proper modifications, while its practical feasibility requires careful consideration.

**Appendix**

**A.1. Analysis of HJB equation**

We mathematically and computationally analyze the HJB equation (6.15). A full analysis of the equation requires a large number of pages, and we only discuss key points that will also be important for other control problems. Here, we assume the case $\omega(t,n)$ is increasing in $n$ for a technical reason, while computationally both increasing and decreasing $\omega(t,n)$ has been computed as demonstrated in the main text. We assume uniform continuity of $S$ for simplicity.

As our problem is a deterministic optimal control problem, we can rely on the dynamic programming principle (Bardi and Dolcetta, 1997). Rigorously, this approach needs a uniform boundedness of the control $q$ such that $0 \leq q_t \leq \bar{q} < +\infty$ with a constant $\bar{q} > 0$. A control problem with an unbounded control variable should be considered because the dynamical system to be controlled is possibly too irregular to ensure the unique existence of a solution to the system (e.g., Cieślak et al. (2021)). We apply this approach to first consider a modified problem with bounded controls and then show that the boundedness is superficial (e.g., Feng et al. (2024)).

For later use, we define viscosity solutions, which are proper weak solutions to the HJB equation. To proceed, collection of all function continuous, lower-semicontinuous, upper-semicontinuous, and smooth on $[0,T] \times [0,+\infty)$ are denoted as $C([0,T] \times [0,+\infty))$, $LSC([0,T] \times [0,+\infty))$, $USC([0,T] \times [0,+\infty))$, and $C^1([0,T] \times [0,+\infty))$, respectively. Viscosity solutions of the modified HJB equation are defined as follows:

***Definition: Viscosity super-solution (resp., sub-solution)***

*A function $\Phi \in LSC([0,T] \times [0,+\infty))$ (resp., $\Phi \in USC([0,T] \times [0,+\infty))$) is said to be a viscosity super-solution (resp., viscosity sub-solutions) if it satisfies the following condition: for any test function $\Psi \in C^1([0,T] \times [0,+\infty))$ such that $\Phi - \Psi$ is minimized (resp., maximized) at $(t,n) = (\hat{t},\hat{n}) \in [0,T] \times [0,+\infty)$, with $\Phi(\hat{t},\hat{n}) = \Psi(\hat{t},\hat{n})$, it follows that*

$$-\frac{\partial \Psi(t,n)}{\partial t} \geq \mathbb{H}\left(t,n,\Phi(t,n),\frac{\partial \Psi(t,n)}{\partial n}\right) \text{ (resp., "}\leq\text{") if } (t,n) = (\hat{t},\hat{n}) \in [0,T) \times (0,+\infty), \quad (6.30)$$

$\Phi(\hat{t},\hat{n}) \geq S(\hat{n})$ *(resp., "$\leq$") if $\hat{t} = T$, and $\Phi(\hat{t},\hat{n}) \geq 0$ (resp., "$\leq$") if $\hat{n} = 0$.*

***Viscosity solution***

*A function $\Phi \in C([0,T] \times [0,+\infty))$ is said to be a viscosity solution if it is uniformly bounded and satisfies the conditions to be a viscosity super-solution and a viscosity sub-solution.*

Viscosity solutions to the modified HJB equations (first and second modified HJB equations) that will appear later can be defined analogously by suitably modifying the Hamiltonian.

By the dynamic programming principle, the HJB equation for this modified problem

$$\Phi(t,n) = \sup_{0 \leq q \leq \bar{q}} J(t,n,q) \tag{6.31}$$

is given as follows (hereafter called modified HJB equation):

$$-\frac{\partial \Phi(t,n)}{\partial t} = \sup_{0 \leq q \leq \bar{q}} \left\{ -\delta \Phi - q \frac{\partial \Phi(t,n)}{\partial n} + 2\sqrt{\omega(t,n)q} - hq \right\}, \quad 0 \leq t < T \text{ and } n > 0 \tag{6.32}$$

and there is no change of boundary and terminal conditions. The (guessed) optimal control in this case clearly satisfies $0 \leq q_t^* \leq \bar{q}$ by the definition of the right-hand side of (6.32). According to Section 1 of Chapter III and IV of Bardi and Dolcetta (1997), the dynamical system (6.8) admits a unique smooth solution for any admissible control $q$ satisfying the bound $0 \leq q_t \leq \bar{q}$ as long as $N_t > 0$, and the value function (6.14) under this setting is a viscosity solution to the modified HJB equation (6.32) (e.g., Barles (2013)). Moreover, the first modified HJB equation admits at most one viscosity solution because of a comparison principle (a viscosity super-solution is not smaller than any viscosity sub-solution; see Theorem 3.7 in Chapter II of Bardi and Dolcetta (1997)) with a slight modification based on the assumption that its viscosity solutions satisfy the terminal and boundary conditions and that these conditions are comparable. The proof is based on a contradiction argument that there exists a bounded viscosity sub-solution that is smaller than a viscosity super-solution at some point in $[0,T] \times [0,+\infty)$. The boundedness of the viscosity solutions is a reasonable assumption because of the following observations of the value function of the modified control problem:

$$\Phi(n,q) \geq J(t,n,0) = 0 \tag{6.33}$$

and

$$\begin{aligned} \Phi(t,n) &\leq \sup_q \int_t^T 2\left(\omega(s,N_s)q_s\right)^{\frac{1}{2}} ds + S(N_T) \\ &\leq \int_t^T 2\left(\bar{K}\bar{q}\right)^{\frac{1}{2}} ds + \bar{S} \\ &\leq 2T\left(\bar{K}\bar{q}\right)^{\frac{1}{2}} + \bar{S} \\ &< +\infty \end{aligned} \tag{6.34}$$

Namely, the value function satisfies the global bound $0 \leq \Phi \leq 2T\left(\bar{K}\bar{q}\right)^{\frac{1}{2}} + \bar{S}$ in $[0,T] \times [0,+\infty)$.

The unique existence of viscosity solutions to the "modified" HJB equation was resolved, as discussed above. However, imposing the boundedness constraint on the control leads to a different control problem from that considered in the main text. A hint to overcome this issue is the formula (6.19), from which we obtain, if $\dfrac{\partial \Phi(t,N_t)}{\partial n}$ is nonnegative then we obtain the estimate

$$0 \leq q_t^* = \frac{\omega(t, N_t)}{\left(h + \dfrac{\partial \Phi(t, N_t)}{\partial n}\right)^2} \leq \frac{\omega(t, N_t)}{h^2} \leq \frac{\bar{W}}{h^2} < +\infty. \tag{6.35}$$

This implies that the (guessed) optimal control is valued within the fixed interval $\left[0, h^{-2}\bar{W}\right]$, and hence adding a constraint $0 \leq q_t \leq \bar{q}$ is innocuous if (6.35) gives the optimal control. In this case, we also need to show the nonnegativity $\dfrac{\partial \Phi(t, N_t)}{\partial n} \geq 0$ as well. This needs the estimate $\dfrac{\partial \Phi}{\partial n} \geq 0$ in $[0,T] \times [0, +\infty)$. For the value function of the modified control problem, this estimate follows due to

$$\Phi(t, n+\varepsilon) - \Phi(t, n) \geq 0 \quad \text{for any} \quad \varepsilon > 0 \tag{6.36}$$

which intuitively follows from the fact that the admissible set is wider for the larger population case $(t, N_t) = (t, n+\varepsilon)$ than that of $(t, N_t) = (t, n)$ and the fact that $S$ is increasing. More rigorously, we can prove (6.36) as follows:

***Proof of (6.36)***

Fix some $(t, n_i) \in [0, T] \times [0, +\infty)$ ($i = 1, 2$) as the case $t = T$ is trivial. The admissible set of controls $q$ for the case $(t, N_t) = (t, n_i)$ is denoted as $\mathbb{Q}_i$. We assume $n_1 \leq n_2$ without any loss of generality. By (6.8), we have

$$N_s^{(i)} = n_i - \int_t^s q_u \, du \quad \text{for} \quad 0 \leq t \leq s \leq T, \tag{6.37}$$

where $N_\cdot^{(i)}$ is $N_\cdot$ associated with $n_i$. We have $N_s^{(2)} \geq N_s^{(1)}$ for $0 \leq t \leq s \leq T$, and hence $\mathbb{Q}_1 \subset \mathbb{Q}_2$. This combined with the increasing nature of $S$ and $\omega$ with respect to $n$ leads to

$$J(t, n_1, q) = \int_t^T \exp(-\delta(s-t)) \left(\frac{1}{\alpha}\left(\omega(s, N_s^{(1)})\right)^\alpha - hq_s\right) ds + \exp(-\delta(T-t)) S\left(N_T^{(1)}\right)$$
$$\leq \int_t^T \exp(-\delta(s-t)) \left(\frac{1}{\alpha}\left(\omega(s, N_s^{(2)})\right)^\alpha - hq_s\right) ds + \exp(-\delta(T-t)) S\left(N_T^{(2)}\right), \tag{6.38}$$
$$= J(t, n_2, q)$$

and hence

$$\Phi(t, n_1) \leq \sup_{q \in \mathbb{Q}_1} J(t, n_1, q) \leq \sup_{q \in \mathbb{Q}_1} J(t, n_2, q) \leq \sup_{q \in \mathbb{Q}_2} J(t, n_2, q) \leq \Phi(t, n_2), \tag{6.39}$$

which concludes the proof.

□

From (6.36), one formally obtains

$$0 \leq \frac{\Phi(t, n+\varepsilon) - \Phi(t, n)}{\varepsilon} \to \frac{\partial \Phi}{\partial n} \quad \text{as} \quad \varepsilon \to +0 \text{, if it exists.} \tag{6.40}$$

The same applies to the modified control problem discussed above, and hence the unique viscosity solution to the modified HJB equation (6.32) satisfies $\frac{\partial \Phi}{\partial n} \geq 0$ of the partial derivative exists.

The remaining task is to show that the unique viscosity solution to the modified HJB equation (6.32) is identical to that of the original HJB equation (6.15). To see this, assume that $\bar{q} = h^{-2}\bar{W}$, which is allowed because of the arbitrariness of $\bar{q} > 0$. By $\frac{\partial \Phi}{\partial n} \geq 0$, the maximization problem in the right-hand side of (6.32) is solved as

$$\arg\max_{0 \leq q \leq \bar{q}} \left\{ -\delta\Phi - q\frac{\partial \Phi(t,n)}{\partial n} + 2\sqrt{\omega(t,n)q} - hq \right\} = \arg\max_{0 \leq q \leq \bar{q}} \left\{ 2\sqrt{\omega(t,n)q} - \left(h + \frac{\partial \Phi(t,n)}{\partial n}\right)q \right\}$$
$$= \frac{\omega(t,n)}{\left(h + \frac{\partial \Phi}{\partial n}\right)^2}, \quad (6.41)$$

from which we arrive at the formula (6.35) for the first modified HJB equation (6.32). Then, the right-hand side of the first modified HJB equation (6.32) is formally the same with that of the original one (6.15).

In summary, we formulated an optimal control problem where the boundedness $0 \leq q_t \leq \bar{q} = h^{-2}\bar{W}$ of the control $q$ is incurred, and showed that the first modified HJB equation (6.32) as an optimality equation admits a unique viscosity solution. This solution is the optimized objective function of this modified problem, and further that the original and first modified HJB equations share the same form. Therefore, the optimized objective function is a viscosity solution to the original HJB equation as well.

Finally, we show that the original HJB equation (6.15) admits at most one viscosity solution, with which we can show that the optimized objective function is actually the value function of the original control problem, and that it is the uniquely viscous solution to the original HJB equation (6.15) as desired. To show this, we need another technical step to show that the second modified HJB equation.

$$-\frac{\partial \Phi(t,n)}{\partial t} = \hat{\mathbb{H}}\left(t, n, \Phi, \frac{\partial \Phi(t,n)}{\partial t}\right), \quad 0 \leq t < T \text{ and } n > 0 \quad (6.42)$$

having the regularized Hamiltonian given by

$$\hat{\mathbb{H}}(t, n, u, p) = -\delta u + \frac{\omega(t,n)}{h + \max\{p, 0\}} \quad (6.43)$$

admits a uniquely viscous solution. We consider the second modified HJB equation (6.42) because the original Hamiltonian $\mathbb{H}(t, n, p)$ blows up at $p = -h$. This prevents us from applying the comparison argument. The modification (6.43) resolves this issue. The second modified HJB equation admits a unique viscosity solution because of the comparison argument with the help of the Lipschitz continuity of $\hat{\mathbb{H}}$ in the last argument. Then, as we concluded for the first modified and original HJB equations, the optimized objective function of the modified control problem is the unique viscosity solution to the second modified HJB equation (6.42) as well. In addition, the unique viscosity solution is the same for the original and second modified HJB equations. The modification (6.43) is thus not innocuous. Consequently, we conclude that the optimized objective function of the modified control problem equals the value function of the original control problem, and that it is a unique viscosity solution to the original HJB equation (6.15).

*Analysis of finite difference schemes*

We argue that the numerical solutions to these schemes converge to the unique viscosity solution of the HJB equation (6.15) under certain conditions.

**Explicit scheme**

First, we rewrite (6.21) as

$$\Phi_{i-1,j} = \Phi_{i,j} - \delta \Delta t \Phi_{i,j} + \frac{\omega(t_{i-1}, n_j) \Delta t}{h + \frac{\Phi_{i,j} - \Phi_{i,j-1}}{\Delta n}} = H(t_i, n_j, \Phi_{i,j}, \Phi_{i,j-1}). \tag{6.44}$$

The function $H : [0,t] \times [0,+\infty) \times \mathbb{R}^2 \to \mathbb{R}$ is continuous in the first and second arguments, strictly increasing in the third argument for a sufficiently small $\Delta t > 0$, and strictly decreasing in the fourth argument for a sufficiently small $\Delta t > 0$ if $\Phi_{i,j} \geq \Phi_{i,j-1}$. These statements are obtained as follows.

To continue, we consider the modified explicit scheme

$$\Phi_{i-1,j} = \Phi_{i,j} - \delta \Delta t \Phi_{i,j} + \frac{\omega(t_{i-1}, n_j) \Delta t}{h + \max\left\{\frac{\Phi_{i,j} - \Phi_{i,j-1}}{\Delta n}, 0\right\}}. \tag{6.45}$$

This scheme is identical to the original explicit scheme (6.21) if $\frac{\Phi_{i,j} - \Phi_{i,j-1}}{\Delta n} \geq 0$. In the sequel, we assume the Courant-Friedrichs-Lewy (CFL) condition

$$1 - \left(\delta + \frac{h^{-2}\bar{W}}{\Delta n}\right) \Delta t \geq 0, \tag{6.46}$$

indicating that the time increment is sufficiently insignificant compared to the space increment. This kind of stability condition is common in explicit schemes (Baňas et al., 2022; Li and Tourin, 2022; Parkinson et al., 2020).

We show the following inequality for the modified explicit scheme:

$$0 \leq \Phi_{i,j-1} \leq \Phi_{i,j} \quad \text{for all} \quad i = 0, 1, 2, ..., I_t \quad \text{and} \quad j = 1, 2, 3, ..., I_n. \tag{6.47}$$

This inequality corresponds to $\frac{\partial \Phi}{\partial n} \geq 0$ and $\Phi \geq 0$ for the original and modified HJB equations.

*Proof of the inequality (6.47)*

The proof is by an induction argument. We firstly show the following inequalities due to the CFL condition (6.46):

$$\frac{\partial}{\partial \Phi_{i,j}} H\left(t_{i-1}, n_j, \Phi_{i,j}, \Phi_{i,j-1}\right) = 1 - \delta \Delta t - \frac{\omega\left(t_{i-1}, n_j\right)\frac{\Delta t}{\Delta n}}{\left(h + \max\left\{\frac{\Phi_{i,j} - \Phi_{i,j-1}}{\Delta n}, 0\right\}\right)^2} \mathbb{I}\left(\Phi_{i,j} \geq \Phi_{i,j-1}\right)$$

$$\geq 1 - \delta \Delta t - \frac{\overline{W}\frac{\Delta t}{\Delta n}}{h^2} \qquad (6.48)$$

$$\geq 1 - \left(\delta + \frac{h^{-2}\overline{W}}{\Delta n}\right)\Delta t$$

$$\geq 0$$

and

$$\frac{\partial}{\partial \Phi_{i,j-1}} H\left(t_{i-1}, n_j, \Phi_{i,j}, \Phi_{i,j-1}\right) = \frac{\omega\left(t_{i-1}, n_j\right)\frac{\Delta t}{\Delta n}}{\left\{h + \max\left\{\frac{\Phi_{i,j} - \Phi_{i,j-1}}{\Delta n}, 0\right\}\right\}^2} \mathbb{I}\left(\Phi_{i,j} \geq \Phi_{i,j-1}\right) \geq 0. \qquad (6.49)$$

The inequality (6.47) trivially holds true for $i = I_t$. Suppose that it holds true for some $i \in \{1, 2, 3, ..., N_t\}$. We have $\Phi_{i-1,0} = 0$ by the boundary condition, and (we are using the fact that $\omega$ is increasing in $n$)

$$\begin{aligned}\Phi_{i-1,j+1} &= H\left(t_{i-1}, n_{j+1}, \Phi_{i,j+1}, \Phi_{i,j}\right) \\ &\geq H\left(t_{i-1}, n_{j+1}, \Phi_{i,j}, \Phi_{i,j}\right) \Leftarrow \text{Induction } \Phi_{i,j+1} \geq \Phi_{i,j} \\ &\geq H\left(t_{i-1}, n_{j+1}, \Phi_{i,j}, \Phi_{i,j-1}\right) \Leftarrow \text{Induction } \Phi_{i,j} \geq \Phi_{i,j-1} \\ &= \left(1 - \delta \Delta t\right)\Phi_{i,j} + \frac{\omega\left(t_{i-1}, n_{j+1}\right)\Delta t}{h + \max\left\{\frac{\Phi_{i,j} - \Phi_{i,j-1}}{\Delta n}, 0\right\}} \\ &\geq \left(1 - \delta \Delta t\right)\Phi_{i,j} + \frac{\omega\left(t_{i-1}, n_j\right)\Delta t}{h + \max\left\{\frac{\Phi_{i,j} - \Phi_{i,j-1}}{\Delta n}, 0\right\}} \Leftarrow \omega\left(t_{i-1}, n_{j+1}\right) \geq \omega\left(t_{i-1}, n_j\right) \\ &\geq \Phi_{i-1,j}\end{aligned} \qquad (6.50)$$

Hence, $\Phi_{i-1,j+1} \geq \Phi_{i-1,j}$. The proof is completed by induction.

$\square$

Consequently, the modified and original explicit schemes are identical; hence, the numerical solutions of the latter converge to the unique viscosity solution of the HJB equation. Taking "max" in (6.45) is therefore innocuous.

We next show that numerical solutions generated by the explicit scheme is stable (numerical solutions are bounded by constants both from below and above irrespective to $\Delta t, \Delta n$). This is proven as follows. For $i = I_t$, we have $0 \leq \Phi_{i,\cdot} \leq \overline{S}$. Next, assume that there exists a constant $L_i$ such that $0 \leq \Phi_{i,\cdot} \leq L_i$ with $L_{I_t} = \overline{S}$ at some $i \in \{1, 2, 3, ..., I_t\}$. We then have

$$\Phi_{i-1,j} = (1-\delta\Delta t)\Phi_{i,j} + \frac{\omega(t_{i-1},n_j)\Delta t}{h+\max\left\{\dfrac{\Phi_{i,j}-\Phi_{i,j-1}}{\Delta n},0\right\}} \leq (1-\delta\Delta t)L_i + \frac{\overline{W}\Delta t}{h}. \tag{6.51}$$

Hence, if we assume the recursion

$$L_{i-1} = (1-\delta\Delta t)L_i + \frac{\overline{W}\Delta t}{h}, \tag{6.52}$$

then by an induction argument we have $0 \leq \Phi_{i,\cdot} \leq L_i$. The remaining task is to find $L_i$. This is possible by an elementary calculus, and we find

$$L_{t_i-k} = (1-\delta\Delta t)^k \overline{S} + \frac{\overline{W}\Delta t}{h}\sum_{m=0}^{k}(1-\delta\Delta t)^m \tag{6.53}$$

A simpler but cruder upper bound of numerical solutions to the explicit scheme is therefore the constant one given by

$$L_i = \overline{S} + \frac{\overline{W}}{h\delta} < +\infty. \tag{6.54}$$

Consequently, we have

$$0 \leq \Phi_{i,j} \leq \overline{S} + \frac{\overline{W}}{h\delta} < +\infty \quad \text{for all} \quad i=0,1,2,...,I_t \quad \text{and} \quad j=0,1,2,...,I_n. \tag{6.55}$$

The explicit scheme is therefore ***monotone***, ***stable***, and ***consistent*** discretization of the modified HJB equation, and numerical solutions of the scheme converges to the continuous unique viscosity solution of this equation due to Theorem 2.1 in Barles and Souganidis (1991) locally uniformly in the domain $(0,T)\times(0,N_{\max})$. Note that the existence of the solution has already been shown: the value function. In our context, the ***monotonicity*** means that we have $\dfrac{\partial}{\partial \Phi_{i,j}}H \geq 0$ and $\dfrac{\partial}{\partial \Phi_{i-1,j-1}}H \geq 0$, which follows due to assuming the CFL condition. The ***stability*** implies that numerical solutions are uniformly bounded, which directly follows from (6.55). Finally, the ***consistency*** means that, at each point $(t,n)\in(0,T)\times(0,N_{\max})$ and sequences $(t_i,n_j)$ converging to $(t,n)$ as $\Delta t, \Delta n \to +0$, the following holds:

$$\frac{\Phi_{i-1,j}-\Phi_{i,j}}{\Delta t} + \delta\Phi_{i,j} - \frac{\omega(t_{i-1},n_j)}{h+\dfrac{\Phi_{i,j}-\Phi_{i,j-1}}{\Delta n}} \to -\frac{\partial\Phi}{\partial t}+\delta\Phi-\frac{\omega(t,n)}{h+\dfrac{\partial\Phi}{\partial n}} \quad \text{as} \quad \Delta t, \Delta n \to +0 \quad \text{with} \quad \Delta t = O(\Delta n) \tag{6.56}$$

for any functions $\Phi \in C^1(\mathbb{R}^2)$. This immediately follows from the fact that we are using the common one-sided differences $\dfrac{\Phi_{i-1,j}-\Phi_{i,j}}{\Delta t}$ and $\dfrac{\Phi_{i,j}-\Phi_{i,j-1}}{\Delta n}$. Consequently, the explicit scheme generates numerical solutions converging locally uniformly to the value function, the unique viscosity solution to the HJB equation, locally uniformly in $(0,T)\times(0,N_{\max})$. See, also Carrillo et al. (2022) and Sun and Guo (2015).

**Semi-implicit scheme**

We only give a short comment on the semi-implicit scheme because the analysis is almost the same as that for the explicit scheme. We can assume the CFL condition (6.46) here, but what is needed is the weaker one $1 - h^{-2}\overline{W}\frac{\Delta t}{\Delta n} \geq 0$ because of implicitly handling the term $\delta \Phi$. This scheme is consistent because it uses one-sided differences. Moreover, it satisfies the inequality (6.50). The **stability** follows an analogous argument to obtain (6.55). The consistency follows as well due to using common one-sided finite differences. Therefore, the semi-implicit scheme is also convergent.

**Implicit scheme**

Finally, we consider an implicit scheme. Firstly, we rewrite (6.25) as

$$H\left(t_{i-1}, n_j, \Phi_{i-1,j}, \Phi_{i-1,j-1}, \Phi_{i,j}\right) = 0 \tag{6.57}$$

with

$$\begin{aligned}H\left(t_{i-1}, n_j, \Phi_{i-1,j}, \Phi_{i-1,j-1}, \Phi_{i,j}\right) &= \frac{\Phi_{i,j} - \Phi_{i-1,j}}{\Delta t} - \delta \Phi_{i-1,j} + \frac{\omega\left(t_{i-1}, n_j\right)}{h + \frac{\Phi_{i-1,j} - \Phi_{i-1,j-1}}{\Delta n}} \\ &= -\left(\frac{1}{\Delta t} + \delta\right)\Phi_{i-1,j} + \frac{\omega\left(t_{i-1}, n_j\right)}{h + \frac{\Phi_{i-1,j} - \Phi_{i-1,j-1}}{\Delta n}}\end{aligned} \tag{6.58}$$

The function $H : [0,t] \times [0,+\infty) \times \mathbb{R}^3 \to \mathbb{R}$ is continuous in the first and second arguments but is not definable if $h + \frac{\Phi_{i-1,j} - \Phi_{i-1,j-1}}{\Delta n} = 0$. Therefore, this point needs to be verified.

We show that the implicit scheme is well-defined; namely, the implicit scheme satisfies

$$\Phi_{i,j-1} < \Phi_{i,j} + h\Delta n \quad \text{for all} \quad i = 0,1,2,...,I_t \quad \text{and} \quad j = 1,2,3,...,I_n \tag{6.59}$$

This inequality corresponds to $\frac{\partial \Phi}{\partial n} > -h$ for the HJB equation and moreover proves that the right-hand side of (6.25) is well-defined. If (6.59) is true, then in (6.58) we have $h + \frac{\Phi_{i-1,j} - \Phi_{i-1,j-1}}{\Delta n} > 0$, and hence the scheme (6.58) is well-defined at least for its solutions. Moreover, numerical solutions are always nonnegative. Notice that the inequality (6.59) is weaker than the inequality (6.47) for the explicit scheme.

*Proof of the inequality (6.59)*

The inequality is proven as follows based on an induction argument. This inequality immediately follows for $i = I_t$ due to the increasing nature of the function $S$: $\Phi_{I_t,j-1} \leq \Phi_{I_t,j}$ for all $j = 1,2,3,...,I_n$ and hence (6.59). Assume that the inequality (6.59) holds true for some $j \in \{1,2,3,...,I_x\}$. We have the boundary condition $\Phi_{i-1,0} = 0$. For each $i = 0,1,2,...,I_t - 1$ and $j = 0,1,2,...,i_n$, we have

$$2C_A \Phi_{i-1,j} = -C_A h\Delta n + \Phi_{i,j} + C_A \Phi_{i-1,j-1}$$
$$+ \sqrt{\{C_A(h\Delta n - \Phi_{i-1,j-1}) + \Phi_{i,j}\}^2 + 4\omega(t_{i-1}, n_j)\Delta t \Delta n} \qquad (6.60)$$
$$> -C_A h\Delta n + \Phi_{i,j} + C_A \Phi_{i-1,j-1} + |C_A(h\Delta n - \Phi_{i-1,j-1}) + \Phi_{i,j}|$$

There are two possible scenarios. If $C_A(h\Delta n - \Phi_{i-1,j-1}) + \Phi_{i,j} \geq 0$ or equivalently $\Phi_{i,j} \geq C_A \Phi_{i-1,j-1} - C_A h\Delta n$, then we have

$$\begin{aligned}2C_A \Phi_{i-1,j} &> -C_A h\Delta n + \Phi_{i,j} + C_A \Phi_{i-1,j-1} + C_A(h\Delta n - \Phi_{i-1,j-1}) + \Phi_{i,j} \\ &= 2\Phi_{i,j} \\ &\geq 2(C_A \Phi_{i-1,j-1} - C_A h\Delta n)\end{aligned} \qquad (6.61)$$

and hence

$$\Phi_{i-1,j} > \Phi_{i-1,j-1} - h\Delta n. \qquad (6.62)$$

If $C_A(h\Delta n - \Phi_{i-1,j-1}) + \Phi_{i,j} < 0$ or equivalently $C_A h\Delta n + \Phi_{i,j} < C_A \Phi_{i-1,j-1}$, then

$$\begin{aligned}2C_A \Phi_{i-1,j} &> -C_A h\Delta n + \Phi_{i,j} + C_A \Phi_{i-1,j-1} - C_A(h\Delta n - \Phi_{i-1,j-1}) - \Phi_{i,j} \\ &= 2(C_A \Phi_{i-1,j-1} - C_A h\Delta n)\end{aligned}, \qquad (6.63)$$

arriving at (6.62).

□

Next, we show the nonnegativity

$$\Phi_{i,j} \geq 0 \quad \text{for all} \quad i = 0,1,2,...,I_t \quad \text{and} \quad j = 0,1,2,...,I_n. \qquad (6.64)$$

***Proof of the inequality (6.64)***

We used this induction again. Then, as in the **Proof of the inequality (6.59)**, for each $i = 0,1,2,...,I_t - 1$ and $j = 0,1,2,...,i_n$, by (6.61) we have

$$C_A \Phi_{i-1,j} \geq \Phi_{i,j}. \qquad (6.65)$$

If $C_A(h\Delta n - \Phi_{i-1,j-1}) + \Phi_{i,j} < 0$, then by (6.63) we have

$$2C_A \Phi_{i-1,j} \geq 2(C_A \Phi_{i-1,j-1} - C_A h\Delta n) \geq \Phi_{i,j} \qquad (6.66)$$

and hence (6.65). These observations, combined with the boundary condition $\Phi_{i,0} = 0$, prove the inequality (6.64).

□

We demonstrate that the implicit scheme is **monotone**, which can be verified as follows. We write the scheme as

$$\Phi_{i-1,j} = \frac{-C_A h \Delta n + C_A \Phi_{i-1,j-1} + \Phi_{i,j} + \sqrt{\{C_A(h\Delta n - \Phi_{i-1,j-1}) + \Phi_{i,j}\}^2 + 4\omega(t_{i-1},n_j)\Delta t \Delta n}}{2C_A} \quad . \tag{6.67}$$
$$= H(t_{i-1}, n_j, \Phi_{i,j}, \Phi_{i-1,j-1})$$

Then, we have the estimates

$$2C_A \frac{\partial}{\partial \Phi_{i-1,j-1}} H(t_{i-1}, n_j, \Phi_{i,j}, \Phi_{i-1,j-1}) = C_A + \frac{-2C_A\{C_A(h\Delta n - \Phi_{i-1,j-1}) + \Phi_{i,j}\}}{2\sqrt{\{C_A(h\Delta n - \Phi_{i-1,j-1}) + \Phi_{i,j}\}^2 + 4\omega(t_{i-1},n_j)\Delta t \Delta n}}$$

$$= C_A \left\{ 1 - \frac{C_A(h\Delta n - \Phi_{i-1,j-1}) + \Phi_{i,j}}{\sqrt{\{C_A(h\Delta n - \Phi_{i-1,j-1}) + \Phi_{i,j}\}^2 + 4\omega(t_{i-1},n_j)\Delta t \Delta n}} \right\} \tag{6.68}$$

$$\geq 0$$

and

$$2C_A \frac{\partial}{\partial \Phi_{i,j}} H(t_{i-1}, n_j, \Phi_{i,j}, \Phi_{i-1,j-1}) = 1 + \frac{2\{C_A(h\Delta n - \Phi_{i-1,j-1}) + \Phi_{i,j}\}}{2\sqrt{\{C_A(h\Delta n - \Phi_{i-1,j-1}) + \Phi_{i,j}\}^2 + 4\omega(t_{i-1},n_j)\Delta t \Delta n}}$$

$$= 1 + \frac{C_A(h\Delta n - \Phi_{i-1,j-1}) + \Phi_{i,j}}{\sqrt{\{C_A(h\Delta n - \Phi_{i-1,j-1}) + \Phi_{i,j}\}^2 + 4\omega(t_{i-1},n_j)\Delta t \Delta n}} \quad . \tag{6.69}$$

$$\geq 0$$

Therefore, the scheme is unconditionally **monotone**. The monotonicity then shows that the inequality (6.47) also applies to the implicit scheme. Therefore, the estimate (6.59) is rough. Note that there is no restriction of the time increment $\Delta t$ for the implicit scheme. The **consistency** of the scheme is immediate from the definition (6.25) as $\Delta t, \Delta n \to +0$ with $\Delta t = O(\Delta n)$.

For **stability**, for each $i \in \{0,1,2,..,.I_t\}$ set $L_i \geq 0$ such that $L_i \geq \max_j \Phi_{i,j}$. Then, by (6.68) and (6.69), we have

$$2C_A \Phi_{i-1,j} = -C_A h \Delta n + \Phi_{i,j} + C_A \Phi_{i-1,j-1}$$
$$+ \sqrt{\{C_A(h\Delta n - \Phi_{i-1,j-1}) + \Phi_{i,j}\}^2 + 4\omega(t_{i-1},n_j)\Delta t \Delta n} \quad . \tag{6.70}$$
$$\leq -C_A h \Delta n + L_i + C_A L_{i-1} + \sqrt{\{C_A(h\Delta n - L_{i-1}) + L_i\}^2 + 4\overline{W}\Delta t \Delta n}$$

This inequality holds true for all $j \in \{1,2,3,..,.I_n\}$, and hence we obtain

$$2C_A L_{i-1} \leq -C_A h \Delta n + L_i + C_A L_{i-1} + \sqrt{\{C_A(h\Delta n - L_{i-1}) + L_i\}^2 + 4\overline{W}\Delta t \Delta n} \tag{6.71}$$

or equivalently

$$C_A h \Delta n + C_A L_{i-1} - L_i \leq \sqrt{\{C_A(h\Delta n - L_{i-1}) + L_i\}^2 + 4\overline{W}\Delta t \Delta n} \quad . \tag{6.72}$$

Assume that $L_{i-1} \geq L_i$. Then, from (6.72) we obtain

$$\{C_A h \Delta n + C_A L_{i-1} - L_i\}^2 \leq \{C_A h \Delta n - C_A L_{i-1} + L_i\}^2 + 4\overline{W}\Delta t \Delta n, \tag{6.73}$$

which can be rearranged as

$$4C_A\left(C_A L_{i-1} - L_i\right)h\Delta n \leq 4\bar{W}\Delta t \Delta n \tag{6.74}$$

and hence

$$L_{i-1} \leq \frac{L_i}{1+\delta\Delta t} + \left(\frac{1}{1+\delta\Delta t}\right)^2 \frac{\bar{W}\Delta t}{h} \leq \frac{L_i}{1+\delta\Delta t} + \frac{1}{1+\delta\Delta t}\frac{\bar{W}\Delta t}{h}, \tag{6.75}$$

from which we arrive at the uniform upper bound

$$L_{I_t - k} = \left(\frac{1}{1+\delta\Delta t}\right)^k \bar{S} + \frac{\bar{W}\Delta t}{h}\sum_{m=0}^{k}\left(\frac{1}{1+\delta\Delta t}\right)^{2m}$$

$$\begin{pmatrix} \leq \bar{S} + \dfrac{\bar{W}\Delta t}{h}\dfrac{1+\delta\Delta t}{\delta\Delta t} \\ = \bar{S} + \dfrac{\bar{W}}{\delta h}(1+\delta T) \\ < +\infty \end{pmatrix}, \quad k = 0, 1, 2, ..., I_t. \tag{6.76}$$

We compare this bound with that of the explicit scheme (6.53). Consequently, we proved that the implicit scheme is **stable**, and the scheme is therefore **stable**, **monotone**, and consistent, and hence satisfies the three conditions for the convergence of numerical solutions in a viscosity sense.